\documentclass[structabstract]{raa}
\usepackage{graphicx,times}
\usepackage{natbib,amssymb}
\usepackage{epsfig}
\usepackage{float}

\providecommand{\bm}[1]{\mbox{\boldmath$#1$\unboldmath}}

\begin{document}
   \title{The Energy Sources of Superluminous Supernovae}

   \volnopage{ {\bf 2018} Vol.\ {\bf XX} No. {\bf XXX}, 000--000}
   \setcounter{page}{1}

   \author{Shan-Qin Wang \inst{1,2,3}, Ling-Jun Wang\inst{4}, Zi-Gao Dai\inst{1,3}}
   \institute{School of Astronomy and Space Science, Nanjing University, Nanjing 210093, China;
          {\it dzg@nju.edu.cn}\\
          \and
          Guangxi Key Laboratory for Relativistic Astrophysics, School of Physical Science and Technology, Guangxi University, Nanning 530004, China; {\it shanqinwang@gxu.edu.cn}\\
          \and
          Key Laboratory of Modern Astronomy and Astrophysics (Nanjing University), Ministry of Education, China
          \and
          Astroparticle Physics, Institute of High Energy Physics, Chinese Academy of Sciences, Beijing 100049, China}

   \date{Received; accepted}

\abstract {Supernovae (SNe) are the most brilliant optical stellar-class explosions.
Over the past two decades, several optical transient survey projects discovered
more than $\sim 100$ so-called superluminous supernovae (SLSNe) whose peak luminosities and
radiated energy are $\gtrsim 7\times 10^{43}$ erg s$^{-1}$ and $\gtrsim
10^{51}$ erg, at least an order of magnitude larger than that of normal SNe.
According to their optical spectra features, SLSNe have been split into
two broad categories of type I that are hydrogen-deficient and type II that
are hydrogen-rich. Investigating and determining the energy sources of SLSNe
would be of outstanding importance for understanding the stellar evolution
and explosion mechanisms. The energy sources of SLSNe can be determined by
analyzing their light curves (LCs) and spectra. The most
prevailing models accounting for the SLSN LCs are the $^{56}$Ni cascade
decay model, the magnetar spin-down model, the ejecta-CSM interaction model,
and the jet-ejecta interaction model. In this \textit{review}, we present
several energy-source models and their different combinations.
 }
\titlerunning{The Energy Sources of Superluminous Supernovae}
\authorrunning{Shan-Qin Wang, Ling-Jun Wang, Zi-Gao Dai}
\maketitle


\section{Introduction}

\label{sec:Intro}

Supernovae (SNe) are believed to be violent explosions of massive
stars or white dwarfs. The peak luminosities and radiated energies of
normal SNe are $\sim 10^{42}-10^{43}$ erg s$^{-1}$ and $\sim 10^{49}$ erg,
respectively. According to their optical spectra
around the peaks, SNe can be divided into type I whose spectra lack hydrogen
lines and type II whose spectra show hydrogen lines \citep{Min1941,Fil1997}.

Over the past two decades, several sky-survey projects for optical transients
have discovered about 100 ultra-luminous SNe (e.g.,
\citealt{Qui2011,Chom2011,Nich2014,Qui2014,DeCia2018,Lunn2018}) whose peak
luminosities and radiated energies are $%
\gtrsim 7\times 10^{43}$ erg s$^{-1}$ (absolute magnitudes in any band must
be $\lesssim -21$ mag \citep{Gal2012} \footnote{\citet{Gal2018} suggest that
the threshold can be set to be $M_g<-19.8$ mag.})
and $\gtrsim 10^{51}$ erg, respectively. These highly luminous SNe are coined
``superluminous supernovae (SLSNe)" (for reviews focusing on observations,
see \citealt{Gal2012,Gal2018}).

Like normal SNe, SLSNe can be divided into types I (hydrogen-poor) and II (hydrogen-rich).
To date, almost all type I SLSNe are helium-deficient and are therefore type Ic.
The spectra of most of type I SLSNe resemble those of the SNe Ic %
\citep{Pas2010,Gal2012,Inse2013,Nich2016b}, especially those of SNe Ic-BL %
\citep{LiuMod2017}. Most SLSNe II are SLSNe IIn whose spectra have narrow-
and intermediate-width H$\alpha$ emission lines \citep{Smith2007}%
, similar to those of SNe IIn \citep{Sch1990,Sch1996,Fil1997}%
. The prototype SLSN IIn is SN 2006gy \citep{Smith2007}.
So far, only two confirmed SLSNe are of type IIL: SN 2008es \citep{Gez2009,Mil2009}
and SN 2013hx \citep{Inse2018}. The similarity between SLSNe Ic/IIn and SNe
Ic-BL/IIn indicates that SLSNe are likely originated from the
explosions of massive stars since SNe Ic-BL/IIn are believed to be
produced by the explosions of massive stars.

According to the characteristics of their light curves (LCs), most SLSNe I
can be divided into two groups: fast-evolving %
\citep{Qui2011,Inse2013,Nich2014} and slow-evolving ones %
\citep{Gal2009,Nich2013,Nich2016a,Inse2017}. However, the LC behaviors of
SLSNe are rather heterogeneous and some SLSNe can be classified into neither
fast-evolving nor slow-evolving, (e.g., Gaia16apd %
\citealt{Nich2017a,Kan2017,Yan2017}), being transitional objects between these
two types. The LCs of some SLSNe I show double-peaked structure %
\citep{Nich2015,NS2016,Smit+2016,Vre2017}. While the LCs of SLSNe II are
more complicated than that of SLSNe I, all of them do not show double-peaked
structure.

SLSNe tend to explode in low-metallicity dwarf galaxies %
\citep{Young2010,Nei2011,Chen2013,Lunn2014,Lunn2015} and the star formation
rates (SFRs) of the host galaxies of SLSNe are usually high. To date, only
very few SLSNe were found in giant, metal-rich galaxies, e.g., SN 2006gy %
\citep{Smith2007} and SN 2017egm \citep{Nich2017b,Bose2018}.

Determining the energy sources powering the LCs of SLSNe is
be of outstanding importance for understanding the stellar evolution
and explosion mechanisms.
We can conclude that the LCs of most ordinary SNe must be powered by $^{56}$Ni
cascade decay (e.g., %
\citealt{Col1969,Col1980,Arn1982,Cap1997,Val2008,Cha2012,PN2013}), and/or
ionized hydrogen recombination (e.g., \citealt{Pop1993,Des2005,Kas2009}),
and a minor of SNe might be powered by ejecta--circumstellar medium (CSM)
interaction (e.g., \citealt{Che1982,Che1994,Chu1994,Chu2009}), or
neutron-star/magnetar spin-down \citep{Ost1971,Mae2007}. Unlike ordinary SNe,
the energy sources of SLSNe are still elusive and in debate.
To date, the most promising energy-source models accounting for the SLSN
observations are pair instability SN model %
\citep{Bar1967,Rak1967,Heg2002,Heg2003} which is essentially the $^{56}$Ni
cascade decay model but required $^{56}$Ni (($\gtrsim 5M_{\odot }$)) are
significantly larger than that for powering ordinary SNe ($\lesssim
0.6M_{\odot }$), the magnetar model
\citep{Kas2010,Woos2010,Cha2012,Cha2013b,
Inse2013,Chen2015,Wang2015a,Wang2016b,Dai2016}, the ejecta--CSM interaction
model \citep{Che2011,Cha2012,Liu2018}, and the fallback (jet-ejecta interaction) model %
\citep{Dex2013}. All These models suppose that the released high-energy photons
get absorbed and heat the ejecta, eventually becoming
UV--optical--NIR emission. In this \textit{Review}, we describe these
energy-source models and their combinations and discuss their implications for SLSNe.


\section{Single Energy-source Models}

The energy-source models interpreting the unique peak (for single-peaked LCs)
or the second peak (for double-peaked LCs) of SNe (and SLSNe) are mainly
the $^{56}$Ni model, the magnetar model, the ejecta-CSM interaction model,
and the fallback (jet-ejecta interaction) model.
In some cases, the combinations of two or three energy sources must be
taken into account. In this section, we focus on the
single energy-source model based on the semi-analytic descriptions.

\subsection{The $^{56}$Ni Model}

\label{subsec:56Ni}

When massive stars explode as Fe-core core-collapse SNe (CCSNe, %
\citealt{Baa1934,Jan2007,Jan2012}), they launch energetic shocks
which can heat the stellar mantles to a temperature $\gtrsim 5 \times 10^9$ K.
Shock-heated silicon shells would synthesize a great amount of radioactive
elements, e.g., $^{56}$Ni, $^{57}$Ni, $^{44}$Ti, and $^{22}$Na, etc. At
early epochs ($\lesssim 500$ days), the power coming from $^{56}$Ni is
significantly larger than that released by all other elements %
\citep{Lun2001,Sol2002}. Due to the large distance, many SLSNe and luminous
SNe lack late-time photometric observations, the contribution
from $^{57}$Ni, $^{44}$Ti, and $^{22}$Na can be therefore neglected in modeling of
LCs of these SNe and radioactive-powered model are equal to $^{56}$%
Ni-powered model.

\begin{figure}
\begin{center}
\includegraphics[width=14.2cm,angle=0]{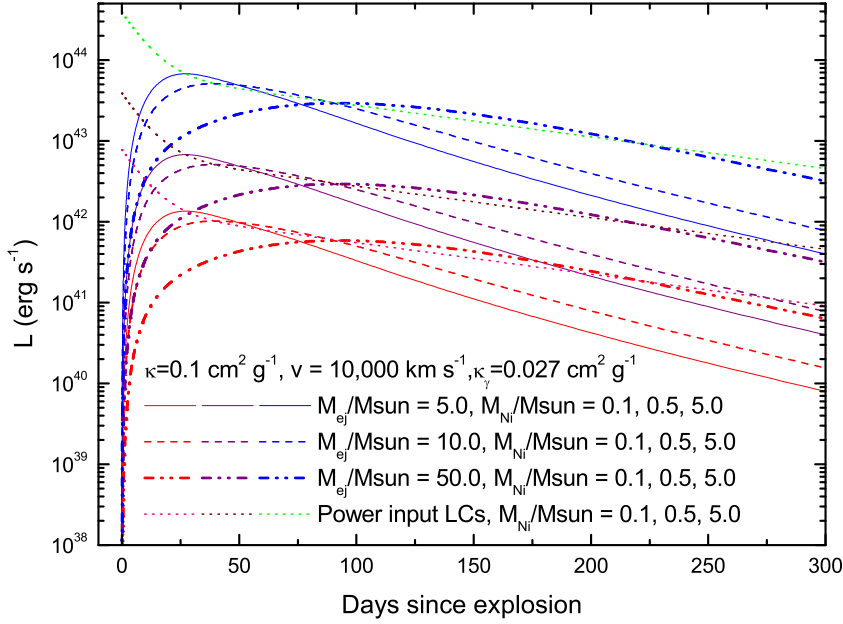}
\caption{LCs powered by $^{56}$Ni cascade decay. We fix $\protect\kappa$ =
0.1 cm$^2$ g$^{-1}$, $v_{\rm sc}$ = $10^9$ cm s$^{-1}$. The LCs powered by same
amount of $^{56}$Ni are presented by same colors.}
\label{fig:ni56model}
\end{center}
\end{figure}

We plot some LCs powered by different amount of $^{56}$Ni in Fig. \ref%
{fig:ni56model}. By fixing $\kappa$ (the optical opacity of the ejecta)
= 0.1 cm$^{2}$ g$^{-1}$, $v_{\rm sc}$ (the scale velocity of the ejecta) = $10^{9}$
cm s$^{-1}$, $\kappa _{\gamma}$ (the gamma opacity of the ejecta)
= 0.027 cm$^{2}$ g$^{-1}$, and setting $M_{%
\mathrm{ej}}$ (the mass of the ejecta) = 5, 10, 50 $M_{\odot }$,
$M_{\mathrm{Ni}}$ (the mass of $^{56}$Ni) = 0.1, 0.5, 5.0 $%
M_{\odot }$, we plot 12 LCs, three of which are $^{56}$Ni cascade decay input
LCs and nine of which are SN LCs powered by $^{56}$Ni cascade decay. Adopting
this set of parameters, Fig. \ref{fig:ni56model} shows that the $^{56}$Ni
model can reasonably explain normal SNe, but is difficult to be used
to explain the LCs of SLSNe ($L_{\mathrm{peak}}\gtrsim 7\times 10^{43}$ erg s$^{-1}$)
since the ratio of $M_{\mathrm{Ni}}$ to $M_{\mathrm{ej}}$ are unreasonably
large (5.0/5.0=1, 5.0/10.0=0.5 for the two most luminous LCs).

Being more luminous than ordinary SNe by a factor of $\sim 10-100$ or more,
the required $^{56}$Ni are usually (significantly) larger
than $\sim 5  M_\odot$ which cannot be synthesized by CCSNe
since the $^{56}$Ni yields of CCSNe cannot exceed $\sim 4 M_\odot$ (\citealt{Ume2008}).
Supposing that the LCs of SLSNe are powered by $^{56}$Ni,
the unique method to solve this problem is supposing that
the explosions are the so-called ``pair instability SNe" (PISNe)
\citep{Bar1967,Rak1967,Heg2002,Heg2003}. For example, \citet{Gal2009}
suggested that SN~2007bi is a PISN; \citet{Cooke2012} studied
two high-redshift SLSNe and concluded that these two SLSNe might be PISNe.

\subsubsection{Fast-Evolving SLSNe~I}

Fast-evolving SLSNe I which constitute a major fraction of
SLSNe~I cannot be explained by the $^{56}$Ni models since the
decline rates of most of them are larger than that of LCs produced
by the $^{56}$Ni model (e.g., \citealt{Qui2011}).
In other words, the $^{56}$Ni masses inferred from the peak luminosities
are significantly larger than that inferred from the late-time light
curves (e.g., \citealt{DeCia2018}).

Moreover, for all fast-evolving SLSNe I, high peak luminosities require
huge amount of $^{56}$Ni while
the narrow LCs indicate that the masses of the ejecta are relatively small.
\citet{Inse2013} and \citet{Nich2014} modeled some fast-evolving SLSNe I
and found that the amount of $^{56}$Ni are $5-30~M_\odot$ and
the masses of the ejecta are between several $M_\odot$ to 30 $M_\odot$, therefore
the ratio of required masses of $^{56}$Ni to the ejecta masses are usually
$\gtrsim$ 50\% or even 100\%, significantly larger than the upper limit
($\sim$ 20\%, \citealt{Ume2008}) of the ratio of the $^{56}$Ni mass to the
ejecta mass.

These studies demonstrated that the $^{56}$Ni models
(including CCSN model and PISN model)
cannot account for fast-evolving SLSNe~I.

\subsubsection{Slow-Evolving SLSNe~I}

Only very few SLSNe I having slow-evolving post-maximum LCs
mimicking that of SNe powered by radioactive elements (mainly $^{56}$Ni)
might be PISNe \citep{Gal2009,Cooke2012} whose $^{56}$Ni masses and
ejecta masses can be $\gtrsim 5M_{\odot }$ and $100-110~M_{\odot }$, respectively.
\citet{Gal2012} proposed that they belong to a distinct class
whose energy source is radioactive elements and named this class ``SLSNe R".

However, \citet{Des2012} argued that SN~2007bi was not
a PISN since its spectra are not consistent with that reproduced by PISN
models. Alternatively, \citet{Des2012} suggested that SN~2007bi was powered by
a magnetar. Moreover, \citet{Inse2017} found that the declined rates of
the LCs at $t-t_{\rm peak}\gtrsim 150$ days of
four slow-evolving SLSNe I (SN~2007bi, PTF12dam, SN~2015bn, and LSQ14an)
are inconsistent with that of LCs reproduced by $^{56}$Co decay, indicating
that they cannot be explained by PISN model since the ejecta masses of
PISNe are very large so that the decline rates of their LCs must be
consistent with that of $^{56}$Co decay rate at $t-t_{\rm peak}\lesssim 500$ days.

\subsubsection{SLSNe~II}

Studies for some type II SLSNe, e.g., SN 2006gy \citep{Agn2009} and CSS121015
\citep{Inse2013}, also demonstrated that the LCs of SLSNe II cannot be explained
by both normal $^{56}$Ni model and the PISN model.

In summary, to date, only a small fraction SLSNe might be powered by the decay of
$^{56}$Ni that were synthesized by PISNe, most SLSNe cannot be explained by $^{56}$Ni
and must be accounted for by other models.

\subsection{The Magnetar Model}

\label{subsec:mag}

CCSN explosions may leave behind fast-rotating neutron stars whose initial
rotational periods ($P_0$) are several milliseconds to several seconds. Based on the
observations for some SN remnants (SNRs), \citet{Ost1971} proposed that
neutron stars with magnetic field strength $B\sim 10^{12}$ G can play a key
role in energizing both SNe and SNRs by injecting their rotational energy to
the ejecta or shells.

The same model has been applied for modeling GRB prompt emission %
\citep{Uso1992,Met2007,Buc2008,Met2011} and GRB afterglows %
(e.g., \citealt{Dai1998a,Dai1998b,Zhang2001,Dai2004,Dai2012}). In these
models, the neutron stars are highly magnetized, $B\sim 10^{14-15}$ G, and
are called ``magnetars".

The magnetar spinning-down model had also been introduced to study SNe. To
account for the LC of SN 2005bf which is not very luminous but cannot be
explained by $^{56}$Ni model, \citet{Mae2007} suggested that the energy
source powering it is a newly-born magnetar and the initial spin period is $%
\sim $ 10 ms. \citet{Woos2010} and \citet{Kas2010} suggested that LCs of SLSNe can be
powered by spinning-down magnetars whose initial spin periods and magnetic strength are $%
\sim 1-5$ ms and $\sim 10^{14-15}$ G, respectively.\footnote{%
\citet{Wang2016a,Wang2017a,Wang2017b} and \citealt{Chen2017} demonstrated
that the LCs of some broad-lined SNe Ic might be powered by millisecond
magnetars if the magnetic strength of these putative magnetars are a few $%
10^{16}$ G.}

\begin{figure}
\begin{center}
\includegraphics[width=14.2cm,angle=0]{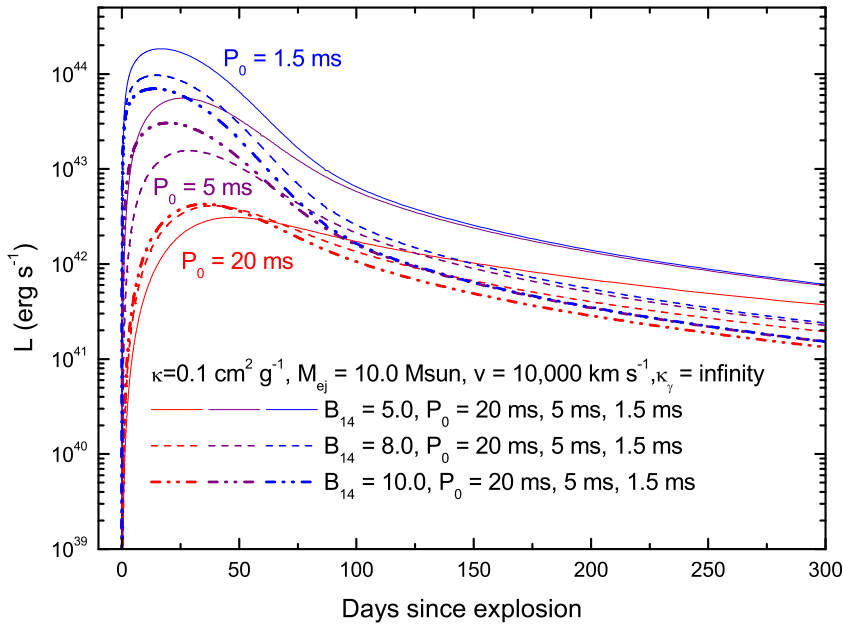}
\caption{LCs powered by magnetar model. We fix $\protect\kappa$ = 0.1 cm$^2$
g$^{-1}$, $M_{\mathrm{ej}} = 10 M_\odot$, $v_{\rm sc}$ = $10^9$ cm s$^{-1}$. The LCs
powered by magnetars with same initial rotational periods are presented by
same colors.}
\label{fig:magmodel}
\end{center}
\end{figure}

The shapes and peak luminosities of the LCs powered by magnetars depend
sensitively on the values of $\kappa $, $M_{\mathrm{ej}}$, $v_{\rm sc}$,
$B$, and $P_0$.
Supposing $\kappa $ = 0.1 cm$^{2}$ g$^{-1}$, $v_{\rm sc}$ = $10^{9}$ cm s$^{-1}$, $%
M_{\mathrm{ej}}=10M_{\odot }$, $\kappa _{\gamma }$ = infinity (full
trapping), and setting $B_{\mathrm{14}}=B/10^{14}\mathrm{G}$ = 5, 8, 10, $%
P_{0}$ = 20, 5, 1.5 ms, we plot 9 LCs powered by magnetars in Figure \ref%
{fig:magmodel}. Adopting this set of parameters, Figure \ref{fig:magmodel}
shows that the magnetar model can reasonably explain normal SNe, luminous
SNe, as well as SLSNe. If we fix the values of $B$
and $P_0$ and vary the
values of $\kappa $, $M_{\mathrm{ej}}$, and $v_{\rm sc}$, we can also get different
LCs. Like the $^{56}$Ni model, larger $M_{\mathrm{ej}}$ and $\kappa$ or
lower $v_{\rm sc}$ would result in dimmer peaks and broad LCs.

\citet{Inse2013}, \citet{Nich2013}, and \citet{Nich2014} used the magnetar
model with full trapping of high energy photons (gamma rays and X-rays)
to fit the LCs of some SLSNe I and found that the LCs reproduced by this model are in
good agreement with the observational data.
As mentioned above, the model adopted by these groups was derived on the assumption of
full-trapping of the gamma-ray and X-ray emission. When the hard emission
was mainly the X-ray emission, this assumption is valid and the LCs
reproduced by this model can be in good/excellent agreement with
observations. If the high-energy emission was dominated by the gamma-ray emission,
a fraction of hard emission would leak from the ejecta before being soften to be UV$-$%
optical$-$IR photons. Therefore, some LCs reproduced by the model with the
assumption have the tails brighter than the observation %
\citep{Nich2014,Chen2015}.

To solve this problem, \citet{Wang2015a} incorporated the leakage effect into
the original magnetar-powered model. If the magnetar emission is
dominated by gamma-ray ($E_{\gamma }\gtrsim 10^{6}$~eV), $\kappa
_{\gamma }\simeq 0.01-0.2$ cm$^{2}$~g$^{-1}$; if the emission is dominated
by X-ray ($10^{2}$~eV $\lesssim E_{\mathrm{X}}\lesssim 10^{6}$~eV), $\kappa
_{\mathrm{X}}\simeq 0.2-10^{4}$ cm$^{2}$~g$^{-1}$, (see Fig. 8 of \citealt{Kot2013}).
By analyzing the late-time LC of PTF12dam, \citet{Chen2015}
also found the magnetar model with full trapping cannot fit the late-time LC
of PTF12dam and introduced a similar trapping factor. Using this revised
magnetar-powered model, the SLSNe whose LC tails cannot be fitted by the
magnetar model with full trapping were well explained, see, e.g., Figure \ref%
{Wang2015a}.

\begin{figure}
\begin{center}
\includegraphics[width=10.2cm,angle=0]{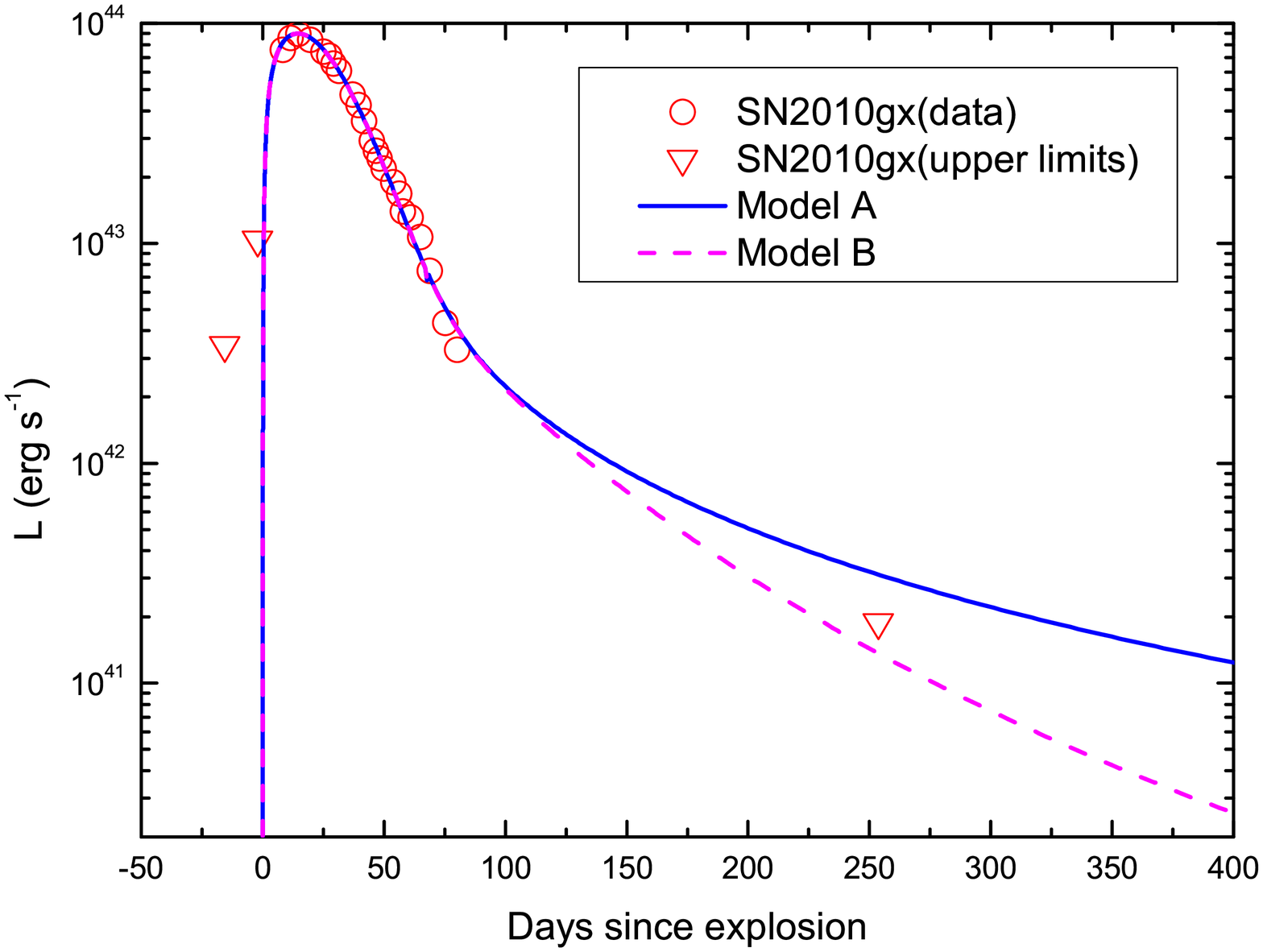}
\includegraphics[width=10.2cm,angle=0]{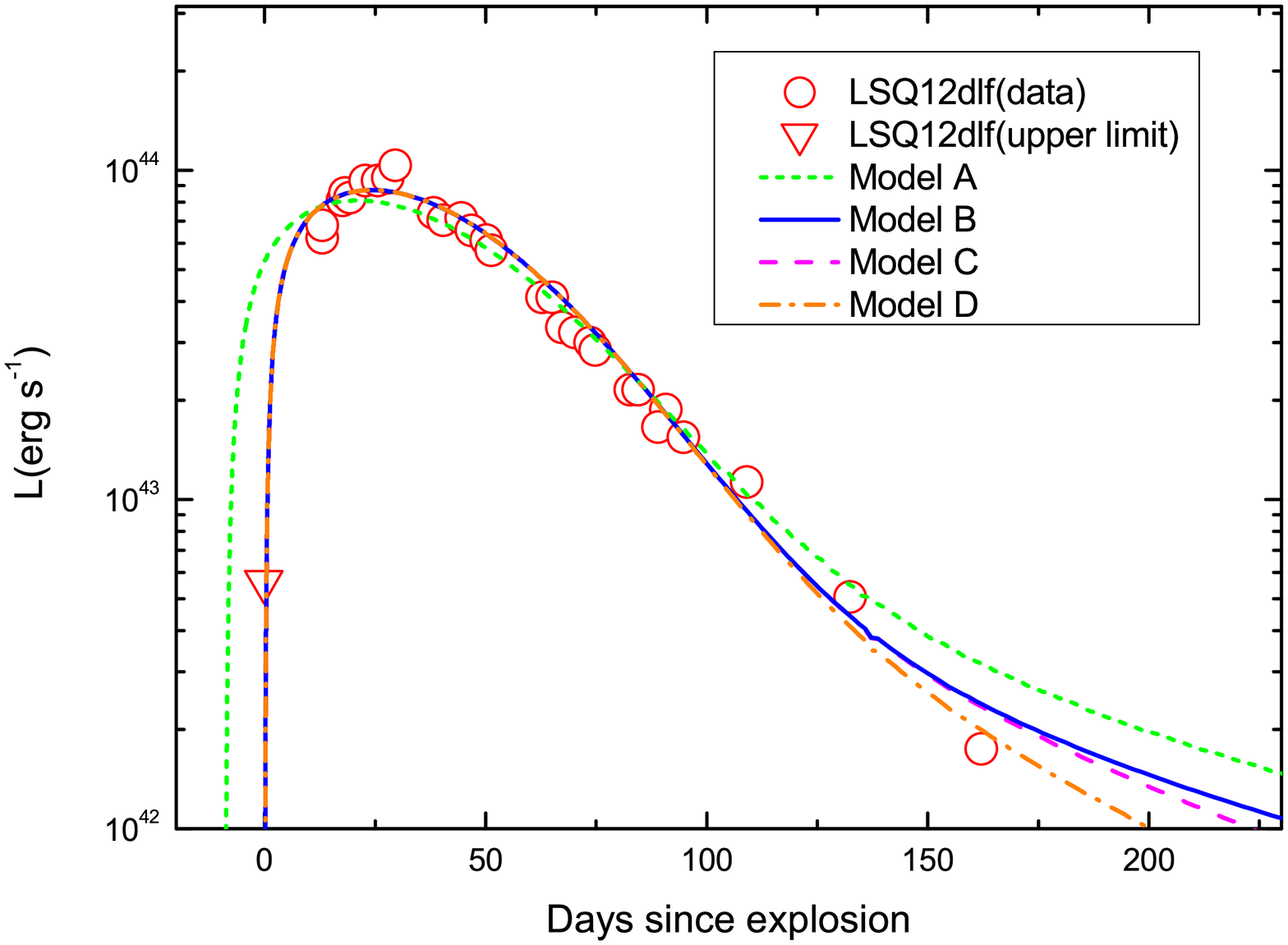}
\caption{LCs in the original magnetar-powered model and the revised
magnetar-powered model for SN~2010gx, and LSQ12dlf. The solid lines and
dashed lines are produced by the magnetar models without and with leakage
effect, respectively \citep{Wang2015a}.}
\label{Wang2015a}
\end{center}
\end{figure}

Although \citet{Woos2010} and \citet{Kas2010} had demonstrated the
acceleration effect is rather notable, the models based on \citet{Arn1982}
all neglect acceleration of the SN
ejecta caused by the magnetar wind. Besides, the photospheric recession effect
is also omitted in these models. \citet{Wang2016b} proposed a new
semi-analytic magnetar-powered model that has taken these two effects into
account. In this new magnetar-powered model, the photospheric velocity of a
SLSN is smaller than the scale velocity $v_{\mathrm{sc}}$ and its evolution
must be fit. Moreover, the scale velocity itself is a running quantity
and is not a parameter or a measurable quantity. Instead, the initial scale
velocity $v_{\mathrm{sc0}}$ is a free parameter.

Using this model, \citet{Liu2017} fitted the data of 19 SLSNe I and found
that the LCs, temperature evolution, and photometric velocity evolution
reproduced by this model are in good agreement with the observations and $%
\sim $ 19$-$97\% of initial rotational energy of the magnetars was converted
to the kinetic energy of the ejecta. Moreover, they found that the initial
kinetic energies of most of these SLSNe are smaller than $\sim 2\times
10^{51}$ erg which is the upper limit of the kinetic energies that can be
provided by neutrino-powered mechanism \citep{Ugl2012,Jan2012,Suk2016}.

\citet{Sok2017} investigated 38 SLSNe I discovered by the Pan-STARRS1 medium
deep survey (PS1 MDS, \citealt{Lunn2018}) and suggested that the SLSNe which
are supposed to be powered by magnetars should be firstly powered by jets
launched from the surfaces of the magnetars. Further investigations for the
magnetar model are needed.

For a SLSN that can be explained by a magnetar, the contribution from $^{56}$%
Ni can be neglected since a SLSN leaving a magnetar is a CCSN whose $^{56}$%
Ni yield is usually rather low, $\lesssim 0.2 M_\odot$, and the luminosity
from this amount of $^{56}$Ni is significantly smaller than that of a SLSN %
\citep{Inse2013}.

\subsection{The Ejecta-CSM Interaction Model}

\label{subsec:inter}

Before the explosions, the progenitors of SNe are surrounded by the
circumstellar winds or material shells ejected from the progenitors just prior
to the SN explosions. After the explosions, the SN ejecta collide with the
winds or shells, generating forward shocks and reverse shocks whose dynamics
can be described by the self-similar solutions \citep{Che1982,Che1994}. In some
extreme cases, the `` pulsational pair-instability (PPI)"
mechanism \citep{Heg2003,Woos2007,Pas2008,Chu2009,Cha+Whe2012} might expel
some shells in different epochs, faster shells might catch up and collide
the slower shells, also generating forward shocks and reverse shocks.
The shock-accelerated electrons emit gamma- and X-ray photons and most of
these photons would be soften to UV$-$optical$-$IR photons. These processes
convert the kinetic energy of the ejecta or the faster shells to the
radiative energy of the SNe and might significantly increase the
luminosities of some SNe if the density of circumstellar wind or shells is
high enough.

The LCs of SLSNe IIn cannot be explained by any model neglecting the
contributions from the interaction-induced shocks. In fact, the ejecta-CSM
interaction model in which the LCs of these SNe are powered by interaction
between the SN ejecta and the hydrogen-rich (and hydrogen-poor) CSM is the
most natural model explaining SNe IIn %
(e.g., \citealt{Chu1994,Mil2010,Zhang2012}), Ibn (e.g., \citealt{Chu2009}), as
well as SLSNe IIn (e.g., \citealt{Smi+McC2007,Mor2013,Nich2014}). Since the
properties of CSM are very complicated, the LCs of luminous SNe IIn, Ibn,
and SLSNe IIn aided by the ejecta-CSM interaction show great complexity
(see \citealt{Smith2016} and references therein).

Many studies have demonstrated that the LCs of SLSNe I and SLSNe IIL whose
spectra are lack of narrow lines indicative of ejecta-CSM interactions or
shell-shell interactions can be explained by the magnetar-powered model.
Although the absence of the interaction signatures in the spectra of SLSNe I and
IIL indicates that the contributions from the interactions can be neglected in
explaining these two classes of SLSNe, the possibility that these SLSNe are
powered by interactions cannot be excluded since the interaction is not
necessary to prompt corresponding signs (e.g., narrow and intermediate-width
H$\alpha$ emission lines).

\citet{Gin2012} used the interaction model to fit the LCs of SN~2010gx (type
I) and SN~2006gy (type IIn). \citet{Nich2014} also used this semi-analytic
model to fit some SLSNe I since the
late-time LCs reproduced by the magnetar-powered model neglecting late-time
leakage are inconsistent with the observations. \citet{Tol2016} argue that
PTF12dam (SLSN I) can be powered by the shell-shell collision.

\begin{figure}
\begin{center}
\includegraphics[width=10.2cm,angle=0]{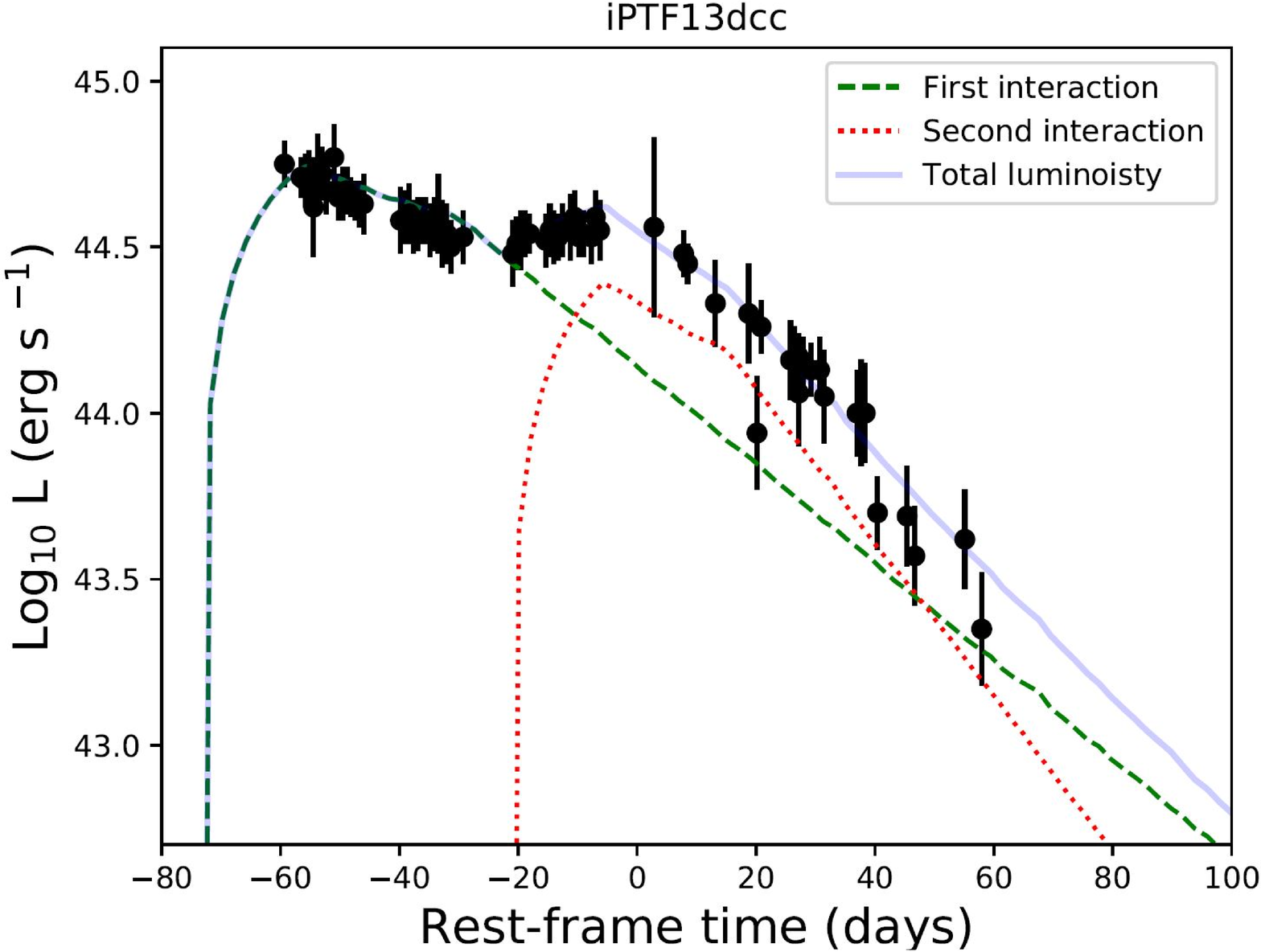}
\includegraphics[width=10.2cm,angle=0]{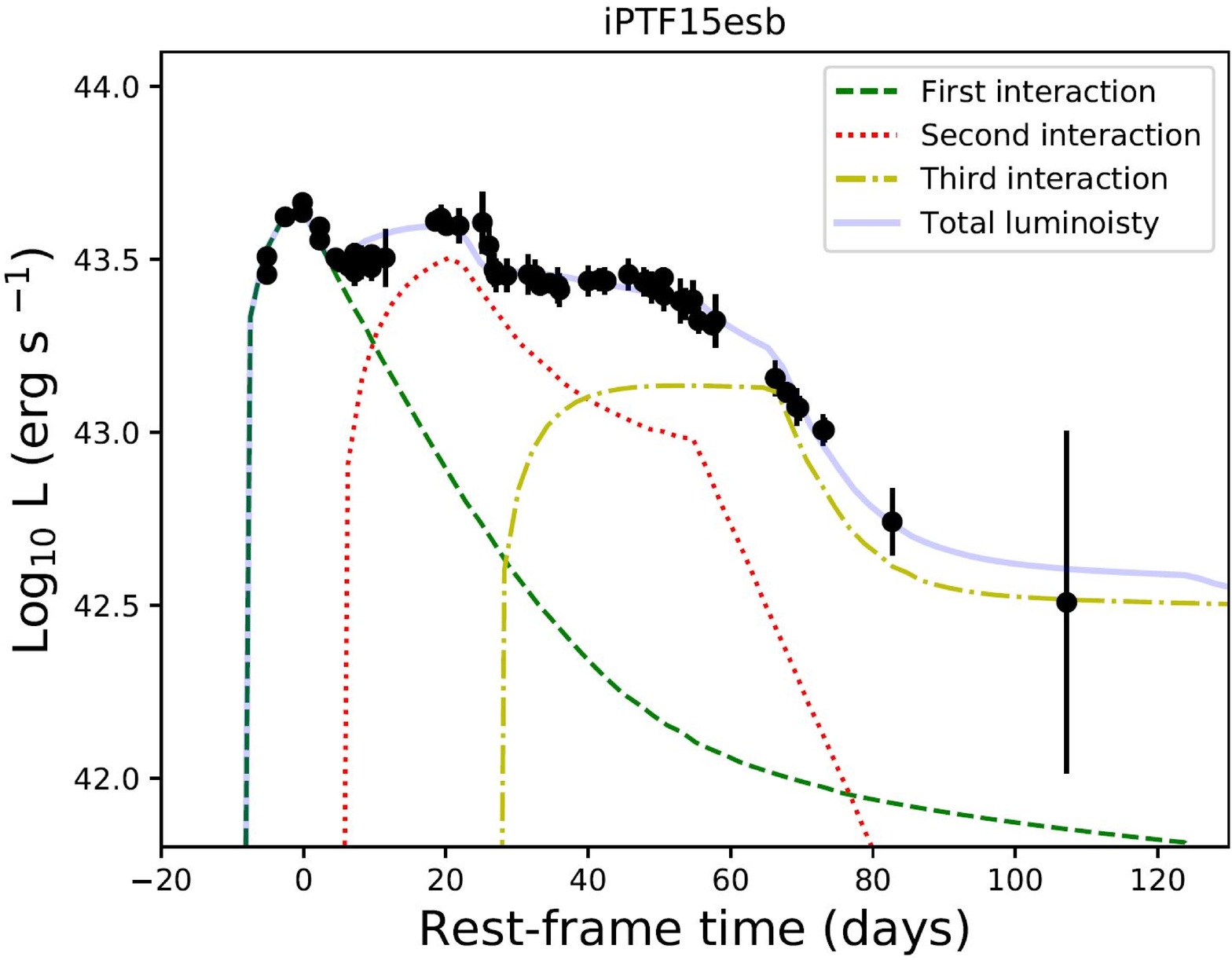}
\caption{The bolometric LCs of iPTF13edcc and iPTF15esb and the LCs
reproduced by the multiple interaction model \citep{Liu2018}.}
\label{fig:Liu2018}
\end{center}
\end{figure}

Recently, \citet{Liu2018} constructed an ejecta--CSM interaction model
involving multiple interactions between the ejecta and different
shells/winds and fit the LCs of iPTF13edcc and iPTF15esb, see
Fig \ref{fig:Liu2018}.

\subsection{The Fallback (jet-ejecta Interaction) Model}

The collapsar model \citep{Woos1993,Mac1999} for gamma ray bursts (GRBs)
proposes that a black hole-disk system can launch a relativistic jet which
can punch a hole in the mantle of a stripped progenitor and produce gamma
ray emission. In this model, the Fe core collapses to a black hole and the
inner mantle material with high angular momentum falls back and forms an
accretion disk.

If the jet cannot breakout and is trapped by the stellar mantle, the energy
carried by the jet will deposited and thermalized to black-body emission. %
\citet{Dex2013} studied this possibility and found that this ``
failed" jet could significantly change the optical LC powered by the
explosion. This model is a jet-ejecta interaction model and can be used to explain SN
LCs with different peak luminosities and durations.
\cite{Wang18} explained the unusual type II-P supernova iPTF14hls which is not
a SLSN as episodic fallback accretion onto a neutron star.
If the deposited energy is large, this model can power a peak
luminosity $\gtrsim 10^{44}$ erg s$^{-1}$ and reproduce the LCs of SLSNe I and II.
\citet{Gao2016} used a similar model to explain ultra-long
GRB 111209A and associated supernova (SN 2011kl).

\citet{Mor2018} used the fallback model to fit 37 SLSNe I and found
that the LCs produced by this model can be consistent with observations.
Moreover, they adopted a typical conversion efficiency $10^{-3}$ and estimated
the required total energy of the accretion disk, finding that the inferred
mass of the accretion disk is $2-700~M_\odot$. They concluded that
only a fraction of SLSNe I whose rising timescales are relative short ($\lesssim 40$ days)
can be explained by this model, or the conversion efficiency must be significantly
larger than $10^{-3}$. As pointed out by \citet{Mor2018}, it is difficult to
distinguish the magnetar model and the fallback model using the LCs produced by these
two models.

\section{Double-energy-source Models}

\subsection{Cooling plus $^{56}$Ni/Magnetar/Interaction models}

\label{subsec:cooling+other}

After the SN explosion, an energetic shock must be launched from the center of
the SN and the shock-breakout (see \citealt{Wax2016} and references therein)
marked by the UV (for non-relativistic shock breakout) or X-ray/gamma-ray
(for relativistic shock breakout) emission would appear and the envelope
would be heated to a temperature of millions of Kelvin (K).
If the progenitors of SNe/SLSNe have extended envelopes, the cooling emission
from shock-heated envelopes would power a LC whose initial luminosity can reach $%
\gtrsim 10^{42}$ erg s$^{-1}$. The cooling emission from a shock-heated
envelope usually peaks at UV$-$optical band and its duration is usually very
short, $\sim $ a few days.

\begin{figure}
\begin{center}
\includegraphics[width=14.2cm,angle=0]{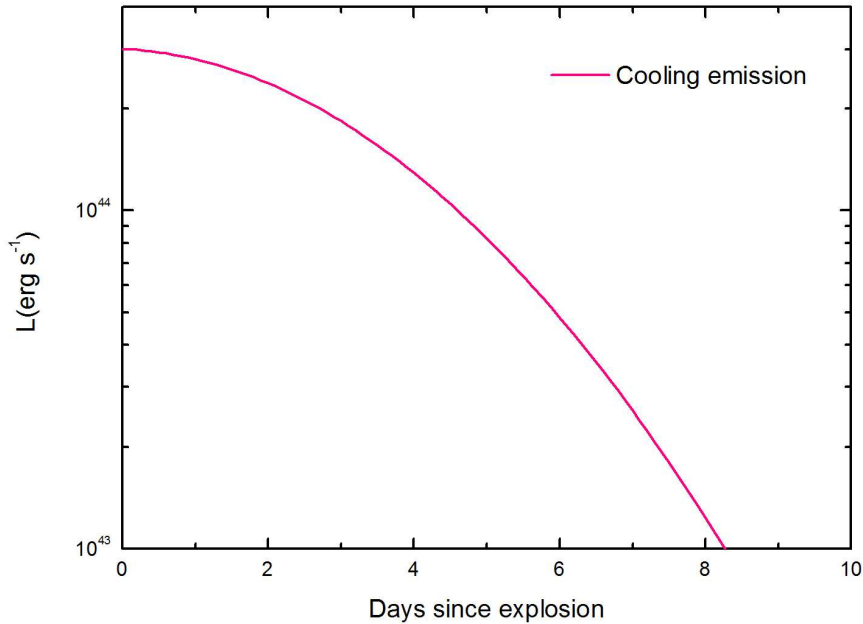}
\caption{LC powered by cooling emission from a shock-heated envelope,
parameters can be found in the text.}
\label{fig:cooling}
\end{center}
\end{figure}

\cite{Piro2015} proposed a concise model that can describe the behavior of
the LC and temperature evolution powered by the cooling emission. The free
parameters of this model are the optical opacity ($\kappa $), mass ($M_{e}$)
and the initial radius ($R_{e}$) of the extended envelope, the mass of the
core of the SN ($M_{c}$), as well as the kinetic energy of the SN ($E_{\mathrm{%
sn}}$). The LC plotted in Fig. \ref{fig:cooling} is yielded by a shock-heated
extended envelope model with $\kappa $ = 0.1 cm$^{2}$ g$^{-1}$, $M_{e}$ = 0.4 $%
M_{\odot }$, $R_{e}$ = 500 $R_{\odot }$, $M_{c}$ = 5 $M_{\odot }$, $E_{%
\mathrm{sn}}=6.75\times 10^{51}$ erg.

\begin{figure}
\begin{center}
\includegraphics[width=14.2cm,angle=0]{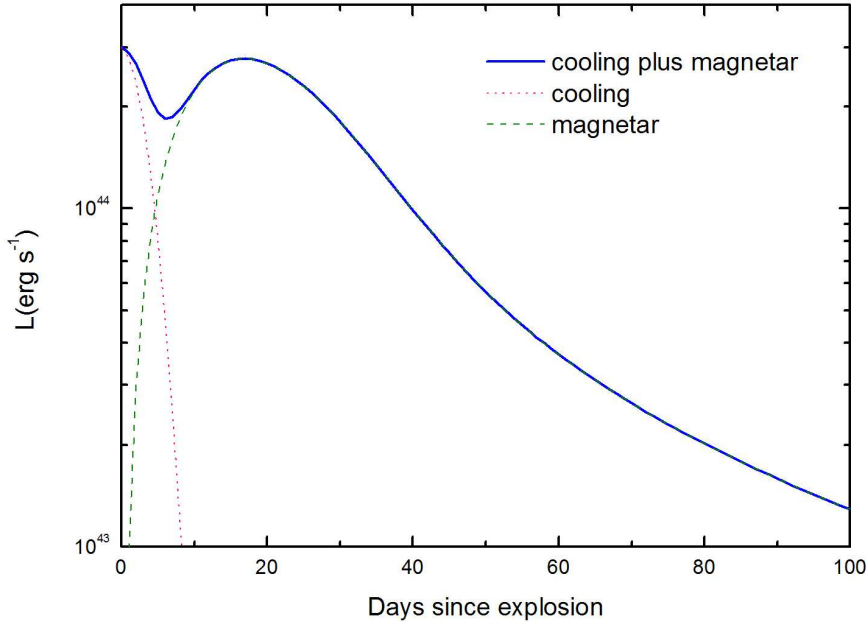}
\caption{The LCs reproduced by the cooling model, the magnetar models, and
the cooling plus magnetar model. $\protect\kappa$ = 0.1 cm$^2$ g$^{-1}$, $%
M_e $ = 0.4 $M_\odot$, $R_e$ = 500 $R_\odot$, $M_c$ = 5 $M_\odot$, $E_{%
\mathrm{sn}} = 6.75 \times 10^{51}$ erg, $v = 1.5 \times 10^9$ cm s$^{-1}$, $%
B = 3\times 10^{14}$ G, $P_0 = 3$ ms.}
\label{fig:cm}
\end{center}
\end{figure}

The SLSNe having double-peaked LCs have been observed, e.g., LSQ14bdq %
\citep{Nich2015}, DES14X3taz \citep{Smit+2016}, PTF12dam \citep{Vre2017},
SN~2006oz \citep{Lel2012, NS2016}, and PS1-10pm \citep{McC2015}. Many groups %
\citep{Nich2015,Smit+2016,Vre2017,NS2016} suggested that the first peaks of
LCs of these SLSNe were powered by the cooling emission from the
shock-heated extended envelopes and the second peaks (main peaks) might be
powered by magnetars or ejecta-CSM interactions.
In this model, the cooling emission power the first peak and $^{56}$Ni synthesized in the
shock-heated ejecta or the magnetar left by the explosion or
the interaction with the ejecta and the CSM
would provide the energy for the second peak and late-time
decay. Energy released by other processes would quickly outshine the cooling emission and
shape the second LC peak. Fig. \ref{fig:cm} shows a LC powered by the
cooling emission and a magnetar.

For the SLSNe whose first peaks were missed by observations or have only
single peaks, cooling emission can be neglected and their whole LCs might be
powered by $^{56}$Ni/magnetar/interaction (see subsections \ref{subsec:56Ni}%
, \ref{subsec:mag}, and \ref{subsec:inter}) or their combinations (see
below).

\subsection{The Magnetar plus $^{56}$Ni Model}

\label{subsec:mag+56Ni}

As mentioned above, the contribution from $^{56}$Ni cascade decay is
significantly smaller than that from other energy sources (magnetar or
interaction) and can be neglected in modeling for SLSNe. However, magnetar
model and interaction model cannot explain Fe lines (if observed) in the
spectra, a moderate amount of $^{56}$Ni is needed for explaining the Fe
lines related to $^{56}$Ni.

Some luminous SNe whose peak magnitudes are between $\sim -20$ mag and $-21$
mag (e.g., \citealt{Deus1995,Schm2000,How2006,San2012,Tad2015,Gre2015,Roy2016,Arc2016,Inse2018})
were also discovered in the past two decades.\footnote{%
Although some authors \citep[e.g.,][]{Ber2016,Inse2018} regarded these
luminous SNe as SLSNe, we still adopt the `` ridgeline" ($M_{%
\mathrm{peak}}=-21$ mag) given by \citet{Gal2012} and suggest that these
luminous SNe belong to a class of `` gap-filler'\ events that
bridge ordinary SNe and SLSNe \citep{Wang2015b,Arc2016}.} \citet{Wang2015b} studied
three luminous SNe Ic-BL and found that they cannot be explained by $^{56}$Ni model.

To solve these two problems, \citet{Wang2015b} proposed that luminous SNe Ic
can also be powered by nascent magnetars whose initial rotational periods ($P_0$)
are $\sim$ 10 ms. Furthermore, \citet{Wang2015b} suggested that the contribution
from some amount of $^{56}$Ni cannot be omitted since the
luminous SNe are not as bright as SLSNe. Therefore, they proposed that these
SNe might be powered by magnetars with $P_0 \sim 10$ ms and $\sim 0.1-0.2 M_\odot
$ of $^{56}$Ni. \citet{Ber2016}, \citet{Met2015} and \citet{Wang2017c} also
employed the double energy sources (magnetar + $^{56}$Ni) to fit the most
luminous GRB-SN, SN~2011kl.

\begin{figure}
\begin{center}
\includegraphics[width=15.2cm,angle=0]{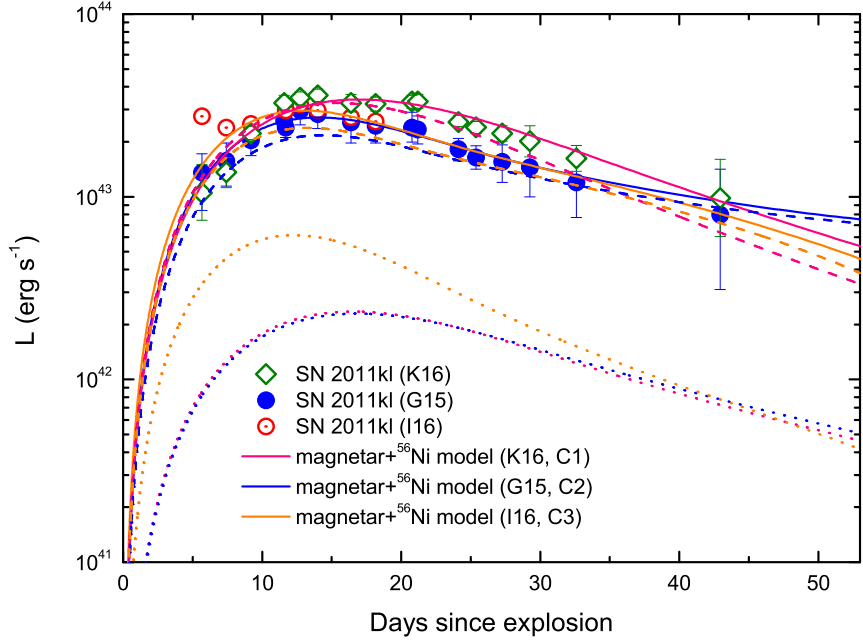}
\caption{The magnetar + $^{56}$Ni model (solid lines) for the LC of
SN~2011kl \citep{Wang2017c}. The dotted lines represent the LCs reproduced
by 0.1 (the dimmer LCs) or 0.2 M$\odot$ (the brighter LCs) of $^{56}$Ni; the
dashed lines represent the LCs powered by magnetars.}
\label{fig:Wang2017c}
\end{center}
\end{figure}

Recently, \citet{Blan2018} studied the multi-band LC and spectra of SN~2017dwh
which is a SLSN I exploded at $z \approx 0.13$. Based on the post-peak
spectra showing a strong absorption line centered near 3200\AA which
is inferred to be Co II and the late-time spectra which also provides
the evidence for the existence of a large mass of Fe-group elements,
\citet{Blan2018} concluded that this SLSN synthesize $\lesssim 0.6 M_\odot$
and used magnetar plus $^{56}$Ni model to model the multi-band LC and
got a rather good result. \citet{Blan2018} found that the best-fitting
parameter of $^{56}$Ni is $0.89^{+0.52}_{-0.58} M_\odot$ (1 $\sigma$ confidence)
whose lower limit ($0.31 M_\odot$) is consistent with the lower
limit ($\lesssim 0.6 M_\odot$) inferred from the spectra.

Based on these studies, we can conclude that some luminous
SNe and SLSNe can be explained by the magnetar plus $^{56}$Ni model.

\subsection{The Interaction plus $^{56}$Ni/Magnetar/Fallback Model}

\label{subsec:int+56Ni}

To fit the LC of SN 2006gy, \citet{Smi+McC2007} constructed a double-energy model
containing the contributions from shock-heated material and $^{56}$Ni
cascade decay. This is the ejecta-CSM interaction plus $^{56}$Ni model. In
this model, the photons coming from shock-heated ejecta and CSM powered the
peak-luminosity as well as the early LC while the late-time LC was powered
by 8 $M_{\odot }$ of $^{56}$Ni. To
synthesize this great amount (8 $M_{\odot }$) of $^{56}$Ni, the explosion
must be a PISN.

\citet{Cha2012} found that the LC of SN~2006gy can be explained by the
ejecta-CSM interaction plus 2 $M_{\odot }$ of $^{56}$Ni. The inferred $^{56}$%
Ni mass is significantly smaller than that inferred by \citet{Smi+McC2007}.
The ejecta mass derived by \citet{Cha2012} is 40 $M_{\odot }$, indicating
that, if this result is correct, SN~2006gy is a CCSN rather than a PISN
since the final masses (ejecta masses) of PISNe must be $\gtrsim
80M_{\odot }$ (see, e.g., \citealt{Cha2013a}). However, how a CCSN can
synthesize 2 $M_{\odot }$ of $^{56}$Ni is still a puzzle. Besides, %
\citealt{Cha2013b} used the interaction plus $^{56}$Ni model to fit 12 SLSNe
(5 SLSNe I and 7 SLSNe II) and got rather good results (they also adopted
other models to fit the LCs of these SLSNe).

To fit the LC of the type Ic supernova iPTF16asu whose
rise time is as short as four days in rest frame, \cite{WangWangXF17b}
constructed a model including early interaction and late-time energy input
from a magnetar. \citet{Chen2018} adopted the interaction plus magnetar model
as well as fallback + interaction model to fit the bolometric
LC of SLSN 2017ens.

\section{Triple-energy-source Model}

\label{subsec:triple-energy-source}

\citet{Yan2015} studied a type I SLSN, iPTF13ehe, and suggested that its nebula
spectra indicate 2.5 $M_{\odot }$ of $^{56}$Ni. If this SLSN was powered by $%
^{56}$Ni, however, the required $^{56}$Ni would be $\gtrsim 13-16 M_{\odot}$,
significantly larger than the value inferred from the spectral analysis.
These facts indicate that the LC of iPTF13ehe cannot be explained by $^{56}$%
Ni model. \citet{Wang2016c} modeled this SLSN using magnetar model and
magnetar plus $^{56}$Ni model and found both these two models can reproduce
the early-time LC of this SLSN.

\begin{figure}
\begin{center}
\includegraphics[width=14.2cm,angle=0]{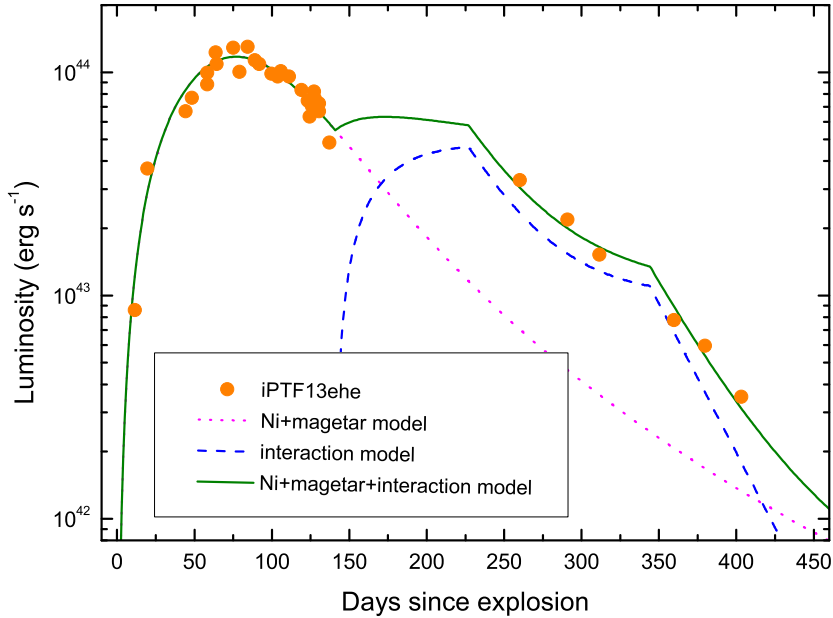}
\caption{Modeling LC of iPTF13ehe using the triple-energy model ($^{56}$Ni +
magnetar + interaction) \citep{Wang2016c}.}
\label{iPTF13ehe}
\end{center}
\end{figure}

Since the late-time spectrum shows narrow H$\alpha $ emission lines indicative
of ejecta-CSM interaction and the late-time LC has a brightening feature, %
\citet{Yan2015} proposed that the ejecta-CSM interaction was triggered when
the ejecta collided with the hydrogen-rich CSM shell expelled prior to the
explosion and produced late-time brightening as well as H$\alpha $ emission lines. %
\citet{Wang2016c} develop a triple energy-source model containing the
contributions from $^{56}$Ni, magnetar and ejecta-CSM interaction to
reproduce the LC of iPTF13ehe; see Fig. \ref{iPTF13ehe}.

\citet{Yan2015} estimated that $\sim $ 15\% of SLSNe I might have late-time H%
$\alpha $ emission lines. Obviously, these SLSNe can be explained by double
(magnetar plus interaction) or triple energy-source model ($^{56}$Ni plus
magnetar plus interaction).

The models containing cooling emission from the shock-heated envelopes of
the SN progenitors and the combinations of $^{56}$Ni+magnetar or $^{56}$%
Ni+interaction or magnetar+interaction are also triple energy source models.
To account for the LC of SN~2011kl which is a luminous type Ic SN, %
\citet{Wang2017c} employed the cooling+$^{56}$Ni+magnetar model.

\section{Discussion}

\label{sec:dis}

\subsection{The Validity of the Models}

Determining the energy sources of the SLSNe is very difficult, but the
energy sources should leave their imprints on the SLSN LCs and spectra. For
example, if a SLSN is powered by a newly-born magnetar, the magnetar wind
would sweep a shell surrounding a bubble. Then the spectra would show
velocity plateau \citet{Kas2010}. If a SLSN is powered by the interaction
between the ejecta and CSM, there might be some narrow emission lines in
their spectra. The narrow H$\alpha $ emission lines are discovered in the
spectra of all SNe IIn and SLSNe IIn, then the interaction model is a valid
model for explaining SNe IIn and SLSNe IIn. However, as pointed out by %
\citet{LiuMod2017}, no narrow emission lines are found in the spectra of 32
SLSNe I collected by them, indicting that these SLSNe might not be powered
by the interaction. Nevertheless, interactions do not prompt emission lines
if the temperature and/or the densities of the CSM is low. Therefore, the
interaction model cannot be ruled out.

To test the one-dimensional magnetar model, \citet{LiuMod2017} analyzed the photospheric
velocity evolutions of 13 SLSNe. They found that only two SLSNe I (PTF10hgi
and PS1-11ap) show slow velocity-evolution feature that is consistent with
the one-dimensional magnetar model and the velocity evolution of the other
11 SLSNe is fast, indicating that the spectra of most SLSNe cannot be
explained by one-dimensional magnetar model. However, the two-dimensional
magnetar model \citep{Chen2016} can destroy the shell structure and the
spectra might quickly evolve.

By studying the spectra of SN~2015bn, %
\citet{Nich2016b} suggest that the strong and relatively narrow O I $\lambda$%
7774 line may indicate the existence of an inner shell swept by a central
engine. Furthermore, they argue that the putative central engine might be a
magnetar, rather than a black hole. Based on 1000 days of photometric
observations for SN~2015bn, \citet{Nich2018} find that the light curve at
very late epoch is consistent with $L\propto t^{-4}$ which can be yielded by
a magnetar spin-down input with inefficient gamma-ray trapping and
pointed out that this light-curve feature indicates the existence of a nascent
magnetar.

\subsection{The Explosion Mechanisms of SLSNe and Luminous SNe}

It is believed that the SNe Ia, Iax, and Ia-CSM might have originated from
the explosions of white dwarfs and other all sub-classes of
normal-luminosity SNe are CCSNe. In contrast, the explosion mechanisms of
SLSNe and luminous SNe are still elusive.

It seems that almost all SLSNe II and fast-evolving SLSNe I ever discovered
might have been originated from the explosions of CCSNe. However, the
explosion mechanisms of slow-evolving SLSNe I are still elusive. %
\citet{Nich2013} modeled the LC of rapid-rising, slow-declining SLSN
PTF12dam and analyzed its spectra, concluding that this SLSN cannot be
explained by the $^{56}$Ni-powered model since the rise time of the LC
produced by the $^{56}$Ni synthesized by PISNe is larger than that of
observed LC. It should be mentioned that
\citet{Koz+Bli2015} proposed a PISN model in which $^{56}$%
Ni is strongly mixed into ejecta and the rise time of LCs produced by this
PISN model is short enough to fit the observational data. As pointed by %
\citet{Mor2017}, however, strong $^{56}$Ni mixing has not yet been
discovered by multidimensional PISN simulations.

\subsection{The Progenitors of SLSNe}

In the past decades, the progenitors of dozens of SNe have been
confirmed directly (see \citealt{Sma2009} and references therein).
Unfortunately, none of these SNe is SLSN since all progenitors of SLSNe are
too distant ($z\gtrsim 0.1$) to be detected before their explosions.

However, many lines of evidence indicate that the progenitors of SLSNe might
be massive stars. The first evidence is that the explosions of white dwarfs
cannot produce so bright transients even if the white dwarfs are
``super-Chandrasekar" ones. The second evidence is that the
spectra of most SLSNe resemble normal-luminosity CCSNe produced by the
explosions of massive stars. The third evidence is that most SLSNe are
located in star-forming dwarf galaxies.\footnote{%
For example, the SFR of the host galaxy of PS1-10bzj is $2-3M_{\odot }$ yr$%
^{-1}$. Since the mass of this host galaxy is $\approx 2.4\times
10^{7}M_{\odot }$, its specific SFR (sSFR) is $\approx 10^{-7}$ yr$^{-1}$ = $%
10^{2}$ Gyr$^{-1}$ \citep{Lunn2013}. Using the high angular-resolution UV
imaging obtained by HST, \citet{Lunn2015} studied the morphological
properties, sizes, and SFR densities of the host galaxies of 16 SLSNe I and
found that these galaxies are compact, irregular galaxies and their
UV-derived SFR densities are high (the averaged value is $\simeq 0.1M_{\odot
}$ yr$^{-1}$ kpc$^{-2}$, \citealt{Lunn2015}).}

Almost all SLSNe are located in the low-metallicity dwarf galaxies,\footnote{%
For example, the metallicity of the host galaxies of SN 2007bi, PS1-10bzj
and SN 2010gx are 1/3 $Z_{\odot }$~\citep{Young2010}, 0.1 $Z_{\odot }$~%
\citep{Lunn2013} and 0.06 $Z_{\odot }$ \citep{Chen2013}, respectively.}
suggesting that the metallicity of the progenitors of SLSNe are rather low.
A massive star with low-metallicity has very low mass-loss rate. However,
the progenitor of a SLSN I must had lost its hydrogen envelope or even the
helium envelope. Therefore, it can be expected that the progenitors of SLSNe I
might be in binary systems and the envelopes of the progenitors must be stripped by
their companions. This mechanism involving mass transfer has another advantage that the
striping process can spun-up the progenitors and is beneficial for the
formations of the millisecond magnetars which might power the LCs of SLSNe.

\subsection{SLSN-GRB Connection}

A minor fraction of SNe Ic have spectra with very broad absorption troughs
which indicate very large photospheric velocities and are named of Broad-lined
SNe Ic (SNe Ic-BL) \citep{Woo2006}. Some SNe Ic-BL associated with long GRBs
has been discovered just after the detections of the corresponding GRBs (%
\citealt{Gal1998,Sta2003,Hjo2003,Gal2004,Camp2006,Maz2006,Ber2011,Sta2011,Mel2012,Xu2013,
 Cano2014,Cano2015,D'Elia2015,Toy2016,Asha2017,Cano2017a}; see %
\citealt{Woo2006,Hjo2012,Cano2017b} for reviews).

The SNe associated with GRBs are dimmer than SLSNe. To date, the most
luminous GRB-SN might be SN~2011kl whose peak bolometric magnitude is $%
\simeq -20.25\pm 0.06$ mag \citep{Kann2016} while the peak
bolometric magnitudes of SLSNe are  $\lesssim
-21$ mag. However, the nature of SN~2011kl is still in debate. %
\citet{Gre2015} suggest that it is an SN while \citet{Ioka2016} argue that
it might be a TDE. If SN 2011kl is a TDE, the peak bolometric magnitude of
most luminous GRB-SN is $\gtrsim -19$ mag.

\citet{Mats2016} investigated the model supposing that the jet successfully
breaks out and generates a GRB while the forward shock and the reverse shock
produced will shock the envelope material and form a hot cocoon. %
\citet{Mats2016} calculated the cocoon emission associated with the black
hole-disk system produced by supermassive population III stars whose masses are $\sim
10^{5}M_{\odot}$ at high redshift ($z\gtrsim 6$) and found that the jet
cocoons will significantly enhance the optical luminosities of the SNe
associated with the GRBs and predicted that the jet--cocoon emission will
power very luminous SNe whose peak luminosities are $\sim 10^{45-46}$ erg s$%
^{-1}$ (the corresponding peak bolometric magnitudes are $\simeq -24$ mag to
$\simeq -26$ mag) after the cocoon breakout of the envelopes. While these
high-redshift, superluminous GRB-SLSNe have not yet been discovered, %
\citet{Mats2016} expect that they will be detected by upcoming NIR
telescopes.

\subsection{SLSNe vs. tidal disruption events (TDEs)}

Judging whether a superluminous optical transient is a SLSN is rather
challenging. As pointed out by \citet{Qui2013a}, it is very difficult to
distinguish among the active galactic nuclei (AGN), tidal disruption events
(TDEs) and SLSNe even if we have multi-band photometry, spectra and high
resolution images.

For example, \citet{Vinko2015} demonstrated that the very luminous ($L_{%
\mathrm{peak}}> 5 \times 10^{44}$ erg s$^{-1}$) optical transient \textit{%
Dougie} might be a super-Eddington TDE, rather than a SLSN; \citet{Dong2016}
suggested that ASASSN-15lh is the most luminous SN discovered so far while %
\citet{Lel2016} argued that it is a TDE.

Although modeling these luminous optical transients would help to
determine their nature, comprehensive observations for their multi-band LCs
and multi-epoch spectra are also needed.

\section{Conclusions}

\label{sec:con}

In the past two decades, SLSNe whose peak luminosities are $\gtrsim
7\times10^{43}$ erg s$^{-1}$ have been discovered by many sky-survey
telescopes and the efforts to unveil the energy sources and nature of
SLSNe have been done by many groups. The LCs and spectra of SLSNe are rather
heterogeneous, reflecting the diverse physical parameters, e.g., energies,
ejecta masses, ejecta velocities, and some other parameters associated with their
central remnants which might play key roles in powering their LCs. Observing
SLSNe and modeling their LCs offer new opportunities to study the
evolution and explosion mechanisms of massive stars.

In this \textit{review}, we present five single energy-source models which have
been used to explain the LCs of SLSNe (and normal SNe), i.e.,
the $^{56}$Ni cascade decay model, the magnetar model,
the ejecta-CSM interaction model, the fallback (jet-ejecta
interaction) model, the cooling model,
as well as their different combinations.

Unlike normal SNe whose LCs can be reasonably explained by the models
mentioned above, it is much less clear how SLSNe can be powered by these
plausible energy sources. The LCs of normal SNe are
mainly powered by $^{56}$Ni cascade decay and the
role of neutron-star spinning-down and the ejecta-CSM interaction can be
neglected in the modeling for most SNe, except for types IIn, Ibn and
Ia-CSM SNe. On the other hand, accumulating observational data and theoretical
modeling indicate that the $^{56}$Ni model cannot account for most of the
LCs of SLSNe and luminous SNe since this scenario requires a huge amount of $%
^{56}$Ni inside the ejecta to account for their peak luminosities and that
the energy from nascent magnetars or ejecta-CSM interaction can play an
essential role in powering the LCs of a majority of SLSNe and luminous SNe.
But we cannot conclude that these SLSNe must be powered by magnetars or
ejecta-CSM interaction and cannot discriminate between these two models even
if the LCs of SLSNe can be explained by one of or both these two models.
To account for the complicated LCs of some SLSNe, the triple-energy-source model
might be employed.

The cooling emission from shock-heated envelopes of the progenitors would
produce the first peaks of the LCs of some SLSNe having double-peaked LCs
and the emission from $^{56}$Ni or magnetar or ejecta-CSM interactions can
power the second peaks. The first peaks
might usually be missed due to the lack of very early observations for some SLSNe.

An appealing unified scenario is that type I SLSNe, type IIL SLSNe, luminous
SNe Ic and normal SNe Ic are powered by neutron stars plus $^{56}$Ni while
type IIn SLSNe, luminous SNe IIn and normal SNe IIn are powered by the
ejecta-CSM interaction plus $^{56}$Ni. In this scenario, the $^{56}$Ni
masses are roughly constant ($\sim 0.1 M_\odot$), the difference of the
neutron-star properties (the initial rotational periods $P_{0}$, the
magnetic field strength $B$) or the ejecta and CSM properties (the ejecta
mass, the ejecta velocity, the CSM mass, the CSM density profile, and so on)
result in different luminosity and rise/decline rate, while the ejecta
masses and velocities determine the LC width.

Extreme conditions are required if these models are valid for explaining SLSNe.
For SLSNe powered by $^{56}$Ni, a huge amount ($\gtrsim 5M_{\odot}
$) of $^{56}$Ni must be synthesized and the SNe might be PISNe; For SLSNe
powered by magnetars, magnetars with very short initial rotational periods ($%
P_{0}\lesssim 10$~ms) and very strong magnetic field strength ($\gtrsim
10^{13-15}$ G) must be left after explosions and the SNe must be CCSNe; For
SLSNe powered by ejecta-CSM interactions or shell-shell interactions, the
progenitors might be $\eta$ Carinae-like stars and must experience strong
wind loss or (multiple) giant eruptions (just) prior to the explosions and
the final explosions can be CCSNe or PISNe, depending on the $^{56}$Ni
masses required.

Although the most prevailing semi-analytic models can yield LCs that are
in good agreement with the photometric observations, their disadvantages are
obvious: neglecting the time- and space-dependent effect of optical opacity, the mixing
effect, and the two/three-dimensional effect. Besides, some very luminous
optical transients having very bright peak luminosities (peak absolute magnitudes
are $\lesssim -20.5$ mag) and very short rising time scales ($\lesssim 10$ days)
cannot be well explained by any models mentioned above.
More detailed modeling might provide more
useful information and eventually determine their nature and their energy
sources.

Determining the energy sources of SLSNe requires more dedicated observations
and theoretical studies. Radio and X-ray observations for the remnants of some SLSNe
also help us to judge whether or not the LCs of SLSNe are powered by
magnetars or the ejecta-CSM interactions or other complicated models.
New sky-survey programs (the Zwicky Transient Facility (ZTF), \citealt{Law2009})
and the upcoming sky-survey programs (e.g., the Large Synoptic Survey Telescope (LSST),
\citealt{Ive2008,LSST2009}),  would discover
more nearby SLSNe and intense follow-up photometric and spectral
observations for them would shed more light on the nature of these optical
transients. Modeling these SLSNe would help to determine their energy sources.


\begin{acknowledgements}

We thank Weikang Zheng for helpful discussions. This work was supported by the National Key Research and Development Program of China (grant No. 2017YFA0402600) and the National Natural Science Foundation of China (grant No. 11573014 and 11833003).

\end{acknowledgements}

\label{lastpage}

\begin{thebibliography}{}\label{thebibliography}

\bibitem[Agnoletto et al.(2009)]{Agn2009} Agnoletto, I, Benetti, S.,
Cappellaro, E., et al. \apj, 2009, 691, 1348

\bibitem[Arcavi et al.(2016)]{Arc2016} Arcavi, I., Wolf, W. M., Howell, D.
A., et al. 2016, \apj, 819, 35

\bibitem[Arnett(1982)]{Arn1982} Arnett, W. D.\ 1982, \apj, 253, 785

\bibitem[Ashall et al.(2017)]{Asha2017} Ashall, C., Pian, E., Mazzali, P.
A., et al. 2017, arXiv:1702.04339

\bibitem[Baade \& Zwicky(1934)]{Baa1934} Baade, W., \& Zwicky, F. 1934,
Phys. Rev, 46, 76

\bibitem[Barkat et al.(1967)]{Bar1967} Barkat, Z., Rakavy, G., \& Sack, N.
1967, \prl, 18, 379

\bibitem[Berger et~al.(2011)]{Ber2011} Berger, E., Chornock, R., Holmes, T.
R., et al. 2011, \apj, 743, 204

\bibitem[Bersten et al.(2016)]{Ber2016} Bersten, M. C., Benvenuto, O. G.,
Orellana, M., \& Nomoto, K. 2016, ApJL, 817, L8


\bibitem[Blanchard et al.(2018)]{Blan2018}Blanchard, P. K., Nicholl, M.,
Berger, E., et al. 2018, arXiv:1810.11051


\bibitem[Bose et al.(2018)]{Bose2018} Bose, S., Dong, S., Pastorello, A., et
al. 2018, \apj, 853, 57

\bibitem[Bucciantini et al.(2008)]{Buc2008} Bucciantini, N., Quataert, E.,
Arons, J., Metzger, B. D., \& Thompson, T. A. 2008, \mnras, 383, L25

\bibitem[Campana et al.(2006)]{Camp2006} Campana, S., Mangano, V., Blustin,
A. J., et al. 2006, \nat, 442, 1008

\bibitem[Cano et al.(2015)]{Cano2015} Cano, Z., de Ugarte Postigo, A.,
Perley, D. et al. \mnras, 2015, 452, 1535

\bibitem[Cano et al.(2014)]{Cano2014} Cano, Z., de Ugarte Postigo, A.,
Pozanenko, A. et al. 2014, \aap, 568, A19

\bibitem[Cano et al.(2017a)]{Cano2017a} Cano, Z., Izzo, L., de Ugarte
Postigo, A., et al. 2017a, \aap, 605, A107

\bibitem[Cano et al.(2017b)]{Cano2017b} Cano, Z., Wang, S. Q., Dai, Z. G.,
Wu, X. F. 2017b, Advances in Astronomy, 8929054, 1

\bibitem[Cappellaro et al.(1997)]{Cap1997} Cappellaro, E., Mazzali, P. A.,
Benetti, S., et al. 1997, \aap, 328, 203

\bibitem[Chatzopoulos \& Wheeler(2012)]{Cha+Whe2012} Chatzopoulos, E., \&
Wheeler, J. C. 2012, \apj, 760, 154

\bibitem[Chatzopoulos et al.(2012)]{Cha2012} Chatzopoulos, E., Wheeler, J.
C., \& Vinko, J. 2012, \apj, 746, 121

\bibitem[Chatzopoulos et al.(2013a)]{Cha2013a} Chatzopoulos, E., Wheeler, J.
C., \& Sean M. C. 2013a, \apj, 776, 129

\bibitem[Chatzopoulos et al.(2013b)]{Cha2013b} Chatzopoulos, E., Wheeler, J.
C., Vinko, J., Horvath, Z. L., \& Nagy, A. 2013b, \apj, 773, 76

\bibitem[Chen et~al.(2017)]{Chen2017} Chen, K.-J., Moriya, T. J., Woosley,
S., Sukhbold, T., Whalen, D. J., Suwa, Y., \& Bromm, V. 2017, \apj, 839, 85

\bibitem[Chen et al.(2016)]{Chen2016} Chen, K.-J., Woosley, S. E., \&
Sukhbold, T. 2016, \apj, 832, 73

\bibitem[Chen et al.(2018)]{Chen2018} Chen, T.-W., Inserra, C., Fraser, M.
et al. 2018, ApJL, 867, L31

\bibitem[Chen et al.(2013)]{Chen2013} Chen, T.-W., Smartt, S. J., Bresolin,
F., et al. 2013, ApJL, 763, L28

\bibitem[Chen et al.(2015)]{Chen2015} Chen, T.-W., Smartt, S.~J.,
Jerkstrand, A., et al. 2015, \mnras, 452, 1567

\bibitem[Chevalier(1982)]{Che1982} {Chevalier}, R.~A. 1982, \apj, 258, 790

\bibitem[Chevalier \& Fransson(1994)]{Che1994} {Chevalier}, R.~A., \& {%
Fransson}, C. 1994, \apj, 420, 268

\bibitem[Chevalier \& Irwin(2011)]{Che2011} Chevalier, R.~A., \& Irwin,
C.~M. 2011, ApJL, 729, L6

\bibitem[Chugai(2009)]{Chu2009} Chugai, N.~N. 2009, \mnras, 400, 866

\bibitem[Chugai \& Danziger(1994)]{Chu1994} Chugai, N. N., \& Danziger, I.
J. 1994, \mnras, 268, 173

\bibitem[Chomiuk et al.(2011)]{Chom2011} Chomiuk, L., Chornock, R.,
Soderberg, A. M., et al. 2011, \apj, 743, 114

\bibitem[Colgate \& McKee(1969)]{Col1969} Colgate, S.~A., \& McKee, C. 1969, %
\apj, 157, 623

\bibitem[Colgate et al.(1980)]{Col1980} Colgate, S.~A., Petschek, A.~G., \&
Kriese, J.~T. 1980, ApJL, 237, L81

\bibitem[Cooke et al.(2012)]{Cooke2012}Cooke, J., Sullivan, M., Gal-Yam, A.,
et al. 2012, \nat, 491, 228

\bibitem[D'Elia et al.(2015)]{D'Elia2015} D'Elia, V., Pian, E., Melandri, A.
et al. 2015, \aap, 577, 116

\bibitem[Dai \& Lu(1998a)]{Dai1998a} Dai, Z. G., \& Lu, T. 1998a, \aap, 333,
L87

\bibitem[Dai \& Lu(1998b)]{Dai1998b} Dai, Z. G., \& Lu, T. 1998b, PhRvL, 81,
4301

\bibitem[Dai(2004)]{Dai2004} Dai, Z. G. 2004, \apj, 606, 1000

\bibitem[Dai \& Liu(2012)]{Dai2012} Dai, Z. G., \& Liu, R. Y. 2012, \apj,
759, 58

\bibitem[Dai et al.(2016)]{Dai2016} Dai, Z. G., Wang, S. Q., Wang, J. S.,
Wang, L. J., \& Yu, Y. W. 2016, \apj, 817, 132

\bibitem[De Cia et al.(2018)]{DeCia2018} De Cia, A., Gal-Yam, A., Rubin, A.,
et al. 2018, ApJ, 860, 100


\bibitem[Dessart \& Hillier(2005)]{Des2005} Dessart, L., \& Hillier, D. J.
2005, \aap, 437, 667

\bibitem[Dessart et al.(2012)]{Des2012} Dessart, L., Hillier, D. J.,
Waldman, R., Livne, E., \& Blondin, S. 2012, \mnras, 426, L76

\bibitem[Deustua et al.(1995)]{Deus1995} Deustua, S., Goldhaber, G., Groom,
D., et al. 1995, IAUC, 6270

\bibitem[Dexter \& Kasen(2013)]{Dex2013} Dexter, J., \& Kasen, D. 2013, \apj%
, 772, 30

\bibitem[Dong et al.(2016)]{Dong2016} Dong, S. B., Shappee, B. J., Prieto,
J. L., et al. 2016, Science, 351, 257

\bibitem[Filippenko(1997)]{Fil1997} Filippenko, A. V. 1997, \araa, 35, 309

\bibitem[Galama et al.(1998)]{Gal1998} Galama, T. J., Vreeswijk, P. M., van
Paradijs, J., et al. 1998, \nat, 395, 670

\bibitem[Gal-Yam(2012)]{Gal2012} {Gal-Yam}, A. 2012, Science, 337, 927

\bibitem[Gal-Yam(2018)]{Gal2018} {Gal-Yam}, A. 2018, \araa, arXiv:1812.01428

\bibitem[Gal-Yam et al.(2009)]{Gal2009} Gal-Yam, A., Mazzali, P., Ofek, E.
O., et al. 2009, \nat, 462, 624

\bibitem[Gal-Yam et al.(2004)]{Gal2004} Gal-Yam, A., Moon, D. S., Fox, D.
B., et al. 2004, ApJL, 609, L59

\bibitem[Gao et al.(2016)]{Gao2016} Gao, H., Lei, W. H., You, Z. Q., \& Xie,
W. 2016, \apj, 826, 141

\bibitem[Gezari et al.(2009)]{Gez2009} Gezari, S., Halpern, J.~P., Grupe,
D., et al. 2009, \apj, 690, 1313

\bibitem[Ginzburg \& Balberg(2012)]{Gin2012} Ginzburg, S., \& Balberg, S.
2012, \apj, 757, 178

\bibitem[Greiner et al.(2015)]{Gre2015} Greiner, J., Mazzali, P. A., Kann,
D. A., et al. 2015, \nat, 523, 189

\bibitem[Heger \& Woosley(2002)]{Heg2002} Heger, A., \& Woosley, S. E. 2002, %
\apj, 567, 532

\bibitem[Heger et al.(2003)]{Heg2003} Heger, A., Fryer, C. L., Woosley, S.
E., et al. 2003, \apj, 591, 288

\bibitem[Hjorth \& Bloom(2012)]{Hjo2012} {Hjorth}, J., \& {Bloom}, J. S.
2012, in Chapter 9 in Gamma-Ray Bursts, ed. C. Kouveliotou, R. A. M. J.
Wijers, \& S. Woosley (Cambridge Astrophysics Series, Vol. 51; Cambridge:
Cambridge Univ. Press), 169

\bibitem[Hjorth et al.(2003)]{Hjo2003} Hjorth, J., Sollerman, J., M{\o }%
ller, P., et al. 2003, \nat, 423, 847

\bibitem[Howell et al.(2006)]{How2006} Howell, D. A., Sullivan, M., Nugent,
P. E., et al. 2006, \nat, 443, 308

\bibitem[Inserra et al.(2017)]{Inse2017} Inserra, C., Nicholl, M., Chen
T.-W., et al. 2017, \mnras, 468, 4642

\bibitem[Inserra et al.(2013)]{Inse2013} Inserra, C., Smartt, S. J.,
Jerkstrand, A., et al. 2013, \apj, 770, 128

\bibitem[Inserra et al.(2018)]{Inse2018} Inserra, C., Smartt, S. J., Gall,
E. E. E., et al. 2018, \mnras, 475, 1046

\bibitem[Ioka et al.(2016)]{Ioka2016} Ioka, K., Hotokezaka, K., \& Piran, T.
2016, \apj, 833, 110

\bibitem[Ivezic et al.(2008)]{Ive2008} Ivezic, Z., Tyson, J. A., Abel, B.,
et al. 2008, arXiv:0805.2366

\bibitem[Janka(2012)]{Jan2012} Janka, H.-T. 2012, Annual Review of Nuclear
and Particle Science, 62, 1, 407

\bibitem[Janka et al.(2007)]{Jan2007} Janka, H.-T., Langanke, K., Marek, A.,
Martinez-Pinedo, G., \& M{\"u}ller, B. 2007, Physics Reports, 442, 1-6, 38

\bibitem[Kasen \& Bildsten(2010)]{Kas2010} Kasen, D., \& Bildsten, L. 2010, %
\apj, 717, 245

\bibitem[Kasen \& Woosley(2009)]{Kas2009} Kasen, Daniel, \& Woosley, S. E.
2009, \apj, 703, 2205

\bibitem[Kangas et al.(2017)]{Kan2017} Kangas, T., Blagorodnova, N.,
Mattila, S., et al. 2017, \mnras, 469, 1246

\bibitem[Kann et al.(2016)]{Kann2016} Kann, D. A., Schady, P., Olivares E.,
F., et al. 2016, arXiv:1606.06791

\bibitem[Kotera et al.(2013)]{Kot2013} Kotera, K., Phinney, E. S., \&
Olinto, A. V. 2013, \mnras, 432, 3228

\bibitem[Kozyreva \& Blinnikov(2015)]{Koz+Bli2015} Kozyreva, A., \&
Blinnikov, S. 2015, \mnras, 454, 4357

\bibitem[Law et al.(2009)]{Law2009} Law, N. M., Kulkarni, S. R., Dekany, R.
G., et al. 2009, \pasp, 121, 1395

\bibitem[Leloudas et al.(2012)]{Lel2012} Leloudas, G., Chatzopoulos, E.,
Dilday, B., et al. 2012, \aap, 541, 129

\bibitem[Leloudas et al.(2016)]{Lel2016} Leloudas, G., Fraser, M., Stone, N.
C., et al. 2016, NatAs, 1, 2

\bibitem[Liu et al.(2018)]{Liu2018} Liu, L. D., Wang, L. J., Wang, S. Q., \&
Dai, Z. G. 2018, \apj, 856, 59

\bibitem[Liu et al.(2017)]{Liu2017} Liu, L. D., Wang, S. Q., Wang, L. J.,
Dai, Z. G., Yu, H., \& Peng, Z. K. 2017, \apj, 842, 26

\bibitem[Liu \& Modjaz(2017)]{LiuMod2017} Liu, Y.-Q., \& Modjaz, M. 2017, %
\apj, 845, 85

\bibitem[LSST Science Collaborations et al.(2009)]{LSST2009} LSST Science
Collaborations et al., 2009, arXiv:0912.0201

\bibitem[Lundqvist et al.(2001)]{Lun2001} Lundqvist, P., Kozma, C.,
Sollerman, J., et al. 2001, \aap, 374, 629

\bibitem[Lunnan et al.(2013)]{Lunn2013} Lunnan, R., Chornock, R., Berger,
E., et al. 2013, \apj, 771, 97

\bibitem[Lunnan et al.(2014)]{Lunn2014} Lunnan, R., Chornock, R., Berger,
E., et al. 2014, \apj, 787, 138

\bibitem[Lunnan et al.(2015)]{Lunn2015} Lunnan, R., Chornock, R., Berger,
E., et al. 2015, \apj, 804, 90

\bibitem[Lunnan et al.(2018)]{Lunn2018} Lunnan, R., Chornock, R., Berger,
E., et al. 2018, \apj, 852, 81

\bibitem[MacFadyen \& Woosley(1999)]{Mac1999} {MacFadyen}, A. I., \& {Woosley%
}, S. E. 1999, \apj, 524, 262

\bibitem[Maeda et al.(2007)]{Mae2007} Maeda, K., Tanaka, M., Nomoto, K., et
al. 2007, \apj, 666, 1069

\bibitem[Matsumoto et al.(2016)]{Mats2016} Matsumoto, T., Nakauchi, D.,
Ioka, K., \& Nakamura, T. 2016, \apj, 823, 83

\bibitem[Mazzali et al.(2006)]{Maz2006} {Mazzali}, P. A., Deng, J., Nomoto,
K., et al. 2006, \nat, 442, 1018

\bibitem[McCrum et al.(2015)]{McC2015} McCrum, M., Smartt, S. J., Rest, A.,
et al. 2015, \mnras, 448, 1206

\bibitem[Melandri et al.(2012)]{Mel2012} Melandri, A., Pian, E., Ferrero,
P., et al. 2012, \aap, 547, 82

\bibitem[Metzger et al.(2011)]{Met2011} Metzger, B. D., Giannios, D.,
Thompson, T. A., Bucciantini, N., \& Quataert, E. 2011, \mnras, 413, 2031

\bibitem[Metzger et al.(2015)]{Met2015} Metzger, B. D., Margalit, B., Kasen,
D., \& Quataert, E. 2015, \mnras, 454, 3311

\bibitem[Metzger et al.(2007)]{Met2007} Metzger, B. D., Thompson, T. A., \&
Quataert, E. 2007, \apj, 659, 561

\bibitem[Minkowski(1941)]{Min1941} Minkowski, R. 1941. \pasp, 53, 22

\bibitem[Miller et al.(2009)]{Mil2009} Miller, A. A., Chornock, R., Perley,
D. A., et al. 2009, \apj, 690, 1303

\bibitem[Miller et al.(2010)]{Mil2010} Miller, A. A., Silverman, J. M.,
Butler, N. R., et al. 2010, \mnras, 404, 305

\bibitem[Moriya et al.(2013)]{Mor2013} Moriya, T. J., Blinnikov, S. I.,
Tominaga, N., Yoshida, N., Tanaka, M., Maeda, K., \& Nomoto, K. 2013, \mnras%
, 428, 1020

\bibitem[Moriya et al.(2017)]{Mor2017} Moriya, T. J., Chen, T.-W., \& Lange,
N. 2017, \apj, 835, 177

\bibitem[Moriya et al.(2018)]{Mor2018} Moriya, T.~J., Nicholl, M. \&
Guillochon, J. \apj, 867, 113

\bibitem[Neill et al.(2011)]{Nei2011} Neill, J. D., Sullivan, M., Gal-Yam,
A., et al. 2011, \apj, 727, 15

\bibitem[Nicholl et al.(2016a)]{Nich2016a} Nicholl, M., Berger, E., Smartt,
S. J., et al. 2016a, \apj, 826, 39

\bibitem[Nicholl et al.(2016b)]{Nich2016b} Nicholl, M., Berger, E.,
Margutti, R., et al. 2016b, ApJL, 828, L18

\bibitem[Nicholl et al.(2017a)]{Nich2017a} Nicholl, M., Berger, E.,
Margutti, R., Blanchard, P. K., Milisavljevic, D., Challis, P., Metzger, B.
D., \& Chornock, R. 2017a, ApJL, 835, L8

\bibitem[Nicholl et al.(2017b)]{Nich2017b} Nicholl, M., Berger, E.,
Margutti, R., Blanchard, P.~K., Guillochon, J., Leja, J., \& Chornock, R.
2017b, ApJL, 845, L8

\bibitem[Nicholl et al.(2018)]{Nich2018}Nicholl, M., Blanchard, P. K.,
Berger, E., et al. 2018, ApJL, 866, L24

\bibitem[Nicholl et al.(2014)]{Nich2014} Nicholl, M., Jerkstrand, A.,
Inserra, C., et al. 2014, \mnras, 444, 2096

\bibitem[Nicholl et al.(2013)]{Nich2013} Nicholl, M., Smartt, S. J.,
Jerkstrand, A., et al. 2013, \nat, 502, 346

\bibitem[Nicholl et al.(2015a)]{Nich2015} Nicholl, M., Smartt, S. J.,
Jerkstrand, A., et al. 2015a, ApJL, 807, L18

\bibitem[Nicholl \& Smartt(2016)]{NS2016} Nicholl, M. \& Smartt, S. J. 2016, %
\mnras, 457, L79

\bibitem[Ostriker \& Gunn(1971)]{Ost1971} Ostriker, J. P., \& Gunn, J. E.
1971, ApJL, 164, L95

\bibitem[Pastorello et al.(2008)]{Pas2008} Pastorello, A., Mattila, S.,
Zampieri, L., et al. 2008, \mnras, 389, 113

\bibitem[Pastorello et al.(2010)]{Pas2010} Pastorello, A., Smartt, S. J.,
Botticella, M. T., et al. 2010, ApJL, 724, L16

\bibitem[Piro(2015)]{Piro2015} Piro, A. L. 2015, ApJL, 808, L51

\bibitem[Piro \& Nakar(2013)]{PN2013} Piro, A. L. \& Nakar, E. 2013, \apj,
769, 67

\bibitem[Popov(1993)]{Pop1993} Popov, D. V. 1993, \apj, 414, 712

\bibitem[Quimby(2014)]{Qui2014} Quimby, R. M. 2014, IAUS, 296, 68

\bibitem[Quimby et al.(2011)]{Qui2011} Quimby, R. M., Kulkarni, S. R.,
Kasliwal, M. M., et al. 2011, \nat, 474, 487

\bibitem[Quimby et al.(2013a)]{Qui2013a} Quimby, R. M., Yuan, F., Akerlof,
C., et al. 2013a, \mnras, 431, 912

\bibitem[Rakavy \& Shaviv(1967)]{Rak1967} Rakavy, G., \& Shaviv, G. 1967, %
\apj, 148, 803

\bibitem[Roy et al.(2016)]{Roy2016} Roy, R., Sollerman, J., Silverman, J.
M., et al. 2016, \aap, 596, 67

\bibitem[Sanders et al.(2012)]{San2012} Sanders, N. E., Soderberg, A. M.,
Valenti, S., et al. 2012, \apj, 756, 184

\bibitem[Schlegel et al.(1990)]{Sch1990} Schlegel, E. M. 1990, \mnras, 244,
269

\bibitem[Schlegel et al.(1996)]{Sch1996} Schlegel, E. M. 1996, \aj, 111, 1660

\bibitem[Schmidt et al.(2007)]{Schm2000} Schmidt, B., Tonry, J., Barris, B.,
et al. 2000, IAUC., 7516, 1

\bibitem[Smartt(2009)]{Sma2009} Smartt, S. J. 2009, \araa, 47, 63

\bibitem[Smith et al.(2007)]{Smith2007} Smith, N., Li, W., Foley, R. J., et
al. 2007, \apj, 666, 1116

\bibitem[Smith \& McCray(2007)]{Smi+McC2007} Smith, N., \& McCray, R. 2007,
ApJL, 671, L17

\bibitem[Smith et al.(2016)]{Smit+2016} Smith, M., Sullivan, M., D'Andrea,
C. B., et al. 2016, ApJL, 818, L8

\bibitem[Smith(2016)]{Smith2016} Smith, N. 2016, arXiv:1612.02006

\bibitem[Soker \& Gilkis(2017)]{Sok2017} Soker, N., \& Gilkis, A. 2017, \apj%
, 851, 95

\bibitem[Sollerman et al.(2002)]{Sol2002} Sollerman, J., Holland, S. T.,
Challis, P., et al. \aap, 386, 944

\bibitem[Stanek et al.(2003)]{Sta2003} Stanek, K. Z., Matheson, T.,
Garnavich, P. M., et al. 2003, ApJL, 591, L17

\bibitem[Starling et al.(2011)]{Sta2011} Starling, R. L. C., Wiersema, K.,
Levan, A. J., et al. 2011, \mnras, 411, 2792

\bibitem[Sukhbold et al.(2016)]{Suk2016} Sukhbold, T., Ertl, T., Woosley, S.
E., Brown, J. M., \& Janka, H.-T. 2016, \apj, 821, 38

\bibitem[Taddia et al.(2015)]{Tad2015} Taddia, F., Sollerman, J., Leloudas,
G., et al. 2015, \aap, 574, 60

\bibitem[Tolstov et al.(2016)]{Tol2016} Tolstov, A., Nomoto, K., Blinnikov,
S., Sorokina, E., Quimby, R., \& Baklanov, P. 2017, ApJ, 835, 266

\bibitem[Toy et al.(2016)]{Toy2016} Toy, V. L., Cenko, S. B., Silverman, J.
M., et al. 2016, \apj, 818, 79

\bibitem[Ugliano et al.(2012)]{Ugl2012} Ugliano, M., Janka, H.-T., Marek,
A., et al. 2012, \apj, 757, 69

\bibitem[Umeda \& Nomoto(2008)]{Ume2008} {Umeda}, H., \& {Nomoto}, K. 2008, %
\apj, 673, 1014

\bibitem[Usov(1992)]{Uso1992} {Usov}, V. V. 1992, Natur, 357, 472

\bibitem[Valenti et al.(2008)]{Val2008} Valenti, S., Benetti, S.,
Cappellaro, E., et al. 2008, \mnras, 383, 1485

\bibitem[Vinko et al.(2015)]{Vinko2015} Vinko, J., Quimby, R. M., Wheeler,
J. C., Ramirez-Ruiz, E., Guillochon, J., Chatzopoulos, E., Marion, G. H., \&
Akerlof, C. 2015, \apj, 798, 12

\bibitem[Vreeswijk et al.(2017)]{Vre2017} Vreeswijk, P. M., Leloudas, G.,
Gal-Yam, A., et al. 2017, \apj, 835, 58

\bibitem[Wang et al.(2017a)]{Wang2017a} Wang, L. J., Cano, Z., Wang, S. Q.,
et al. 2017a, \apj, 851, 54

\bibitem[Wang et al.(2016a)]{Wang2016a} Wang, L. J., Han, Y. H., Xu, D., et
al. 2016a, \apj, 831, 41

\bibitem[Wang et al.(2016b)]{Wang2016b} Wang, L. J., Wang, S. Q., Dai, Z.
G., Xu, D., Han, Y. H., Wu, X. F., \& Wei, J. Y. 2016b, \apj, 821, 22

\bibitem[Wang et al.(2017b)]{WangWangXF17b} Wang, L. J., Wang, X. F., Cano,
Z., et al. 2017b, submitted (arXiv:1712.07359)

\bibitem[Wang et al.(2018)]{Wang18} Wang, L. J., Wang, X. F., Wang, S. Q.,
et al. 2018, \apj, 865, 95

\bibitem[Wang et al.(2017c)]{Wang2017b} Wang, L. J., Yu, H., Liu, L. D.,
Wang, S. Q., Han, Y. H., Xu, D., Dai, Z. G., Qiu, Y. L., \& Wei, J. Y.
2017c, \apj, 837, 128

\bibitem[Wang et al.(2017d)]{Wang2017c} Wang, S. Q., Cano, Z., Wang, L. J.,
et al. 2017d, \apj, 850, 148

\bibitem[Wang et al.(2016c)]{Wang2016c} Wang, S. Q., Liu, L. D., Dai, Z. G.,
Wang, L. J., \& Wu, X. F. 2016c, \apj, 828, 87

\bibitem[Wang et al.(2015a)]{Wang2015a} Wang, S. Q., Wang, L. J., Dai, Z.
G., \& Wu, X. F. 2015a, \apj, 799, 107

\bibitem[Wang et al.(2015b)]{Wang2015b} Wang, S. Q., Wang, L. J., Dai, Z.
G., \& Wu, X. F. 2015b, \apj, 807, 147

\bibitem[Waxman \& Katz(2016)]{Wax2016} Waxman, E., \& Katz, B. 2016,
arxiv:1607.01293

\bibitem[Woosley(1993)]{Woos1993} Woosley, S. E. 1993, \apj, 405, 273

\bibitem[Woosley(2010)]{Woos2010} Woosley, S. E. 2010, ApJL, 719, L204

\bibitem[Woosley et al.(2007)]{Woos2007} Woosley, S. E., Blinnikov, S., \&
Heger, A. 2007, \nat, 450, 390

\bibitem[Woosley \& Bloom(2006)]{Woo2006} Woosley, S. E., \& Bloom, J. S.
2006, \araa, 44, 507

\bibitem[Xu et al.(2013)]{Xu2013} Xu, D., de Ugarte Postigo, A., Leloudas,
G., et~al. 2013, \apj, 776, 98

\bibitem[Yan et al.(2017)]{Yan2017} Yan, L., Quimby, R., Gal-Yam, A., et al.
2017, \apj, 840, 57

\bibitem[Yan et al.(2015)]{Yan2015} Yan, L., Quimby, R., Ofek, E., et al.
2015, \apj, 814, 108

\bibitem[Young et al.(2010)]{Young2010} Young, D. R., Smartt, S. J.,
Valenti, S., et al. 2010, \aap, 512, A70

\bibitem[Zhang \& M{\'e}sz{\'a}ros(2001)]{Zhang2001} Zhang, B., \& M{\'e}sz{%
\'a}ros, P. 2001, ApJL, 552, L35

\bibitem[Zhang et al.(2012)]{Zhang2012} Zhang, T. M., Wang, X. F., Wu, C.,
et al. 2012, \aj, 144, 131

\end{thebibliography}
\end{document}